\documentclass[a4paper,aps,prd,showpacs,preprintnumbers,twocolumn,nofootinbib,superscriptaddress]{revtex4-1}

\usepackage[english]{babel}

\usepackage[utf8x]{inputenc}
\usepackage[T1]{fontenc}
\usepackage{ucs}
\usepackage{microtype}
\usepackage{newtxtext,newtxmath}
\usepackage{color}

\usepackage{url}
\usepackage{hyperref}

\usepackage{xspace}
\usepackage{lastpage}
\usepackage{stmaryrd}

\usepackage{graphicx}
\usepackage{empheq}
\usepackage{ifthen}
\usepackage{tensor}
\usepackage{paralist}

\usepackage{amsmath}
\usepackage{amsfonts}
\usepackage{amssymb}
\usepackage{mathrsfs}
\usepackage{extarrows}
\usepackage{verbatim}
\usepackage{upgreek}
\usepackage{wasysym}
\usepackage{esint}

\usepackage{natbib}

\newcommand\ph{\ensuremath{\varphi}}
\newcommand\eps{\ensuremath{\varepsilon}}

\newcommand\define{\equiv}

\newcommand\vect[1]{\boldsymbol{#1}}
\newcommand{\mat}[1]{\boldsymbol{#1}}
\newcommand\cplx[1]{\underline{#1}}
\newcommand\ex[1]{\mathrm{e}^{#1}}
\renewcommand\i{\ensuremath{\mathrm{i}}}

\renewcommand\Re{\ensuremath{\mathrm{Re}}}
\renewcommand\Im{\ensuremath{\mathrm{Im}}}
\newcommand{\Arg}{\ensuremath{\mathrm{Arg}}}

\newcommand{\tr}{\mathrm{tr}}

\newcommand\e[1]{_{\text{#1}}}
\newcommand\h[1]{^{\text{#1}}}
\newcommand\U[1]{\:\mathrm{#1}}

\newcommand{\dd}{\mathrm{d}}

\newcommand{\pd}[3][]{\frac{\partial^{#1} #2}{\partial {#3}^{#1}}}
\newcommand{\ddf}[3][]{\frac{\dd^{#1} #2}{\dd {#3}^{#1}}}

\renewcommand\lim[2]{\underset{ #1 \rightarrow #2 }{ \mathrm{lim} } \,}

\newcommand{\delimiters}[4][]{
\ifthenelse{ \equal{#1}{1} }{  #2 #3 #4  }
					{ \ifthenelse{\equal{#1}{2}}{ \big#2 #3 \big#4 }
						{ \ifthenelse{\equal{#1}{3}}{ \Big#2 #3 \Big#4 }
							{ \ifthenelse{\equal{#1}{4}}{ \bigg#2 #3 \bigg#4 }
								{ \ifthenelse{\equal{#1}{5}}{ \Bigg#2 #3 \Bigg#4 }
									{ \left#2 #3 \right#4 }
								}
							}
						}
					}
													}

\newcommand{\pa}[2][]{\delimiters[#1]{(}{#2}{)}}
\newcommand{\pac}[2][]{\delimiters[#1]{[}{#2}{]}}
\newcommand{\paac}[2][]{\delimiters[#1]{\{}{#2}{\}}}
\newcommand{\abs}[2][]{\delimiters[#1]{|}{#2}{|}}
\newcommand{\ev}[2][]{\delimiters[#1]{\langle}{#2}{\rangle}}

\newcommand{\aap}{A{\&}A}

\newcommand{\mnras}{MNRAS}

\newcommand{\jcap}{JCAP}

\newcommand{\source}{\mathcal{S}}
\newcommand{\image}{\mathcal{I}}

\newcommand{\interior}{\mathrm{int}}
\newcommand{\exterior}{\mathrm{ext}}
\newcommand{\cir}{\mathscr{C}}

\newcommand{\lens}{\lambda}
\newcommand{\quadrupole}{\mathcal{Q}}
\newcommand{\moment}{\mathcal{M}}
\newcommand{\kernel}{\mathcal{K}}
\newcommand{\normalization}{N}
\newcommand{\flexionF}{\mathcal{F}}
\newcommand{\flexionG}{\mathcal{G}}
\newcommand{\res}{\mathrm{res}}

\newcommand{\FLUletter}{FLU17\xspace}
\newcommand{\FLUshear}{FLU18a\xspace}

\begin{document}

\title{Weak lensing distortions beyond shear}

\author{Pierre Fleury}
\email{pierre.fleury@unige.ch}
\affiliation{D\'{e}partment de Physique Th\'{e}orique, Universit\'{e} de Gen\`{e}ve,\\
24 quai Ernest-Ansermet, 1211 Gen\`{e}ve 4, Switzerland}

\author{Julien Larena}
\email{julien.larena@uct.ac.za}
\affiliation{Department of Mathematics and Applied Mathematics\\
University of Cape Town,
Rondebosch 7701, South Africa}

\author{Jean-Philippe Uzan}
\email{uzan@iap.fr}
\affiliation{
            Institut d'Astrophysique de Paris, CNRS UMR 7095, Universit\'e Pierre et Marie Curie - Paris VI, 98 bis Boulevard Arago, 75014 Paris, France \\
           Sorbonne Universit\'es, Institut Lagrange de Paris, 98 bis, Boulevard Arago, 75014 Paris, France.}

\begin{abstract}
When a luminous source is extended, its distortions by weak gravitational lensing are richer than a mere combination of magnification and shear. In a recent work, we proposed an elegant formalism based on complex analysis to describe and calculate such distortions. The present article further elaborates this finite-beam approach, and applies it to a realistic cosmological model. In particular, the cosmic correlations of image distortions beyond shear are predicted for the first time. These constitute new weak-lensing observables, sensitive to very-small-scale features of the distribution of matter in the Universe. While the major part of the analysis is performed in the approximation of circular sources, a general method for extending it to noncircular sources is presented and applied to the astrophysically relevant case of elliptic sources.
\end{abstract}

\date{\today}
\maketitle

\section{Introduction}

The standard theory of weak gravitational lensing is built upon a relativistic formalism whereby light beams, and hence their sources, are infinitesimal~\cite{1961RSPSA.264..309S, 1992grle.book.....S}. In this context, gravitation acts on photon beams via tidal forces, which by essence can only produce three classes of effects: convergence, shear, and rotation. In particular, shear, which is a change in the apparent ellipticity of an image, is the only distortion that infinitesimal sources can undergo. The weak shear field and its statistical properties currently represent a key observable in cosmology.

In a previous article, Ref.~\cite{Fleury:2017owg}, hereafter \FLUletter, we argued that the infinitesimal-beam approximation is conceptually incorrect when light propagates through matter, whose distribution always vary on scales that are eventually shorter than the beam's cross-sectional diameter, provided one adopts a sufficient resolution. We addressed this problem by designing a \emph{finite-beam} formalism for weak lensing, which allowed us, in particular, to solve the so-called Ricci-Weyl dichotomy~\cite{1964SvA.....8...13Z, 1966RSPSA.294..195B, 1981GReGr..13.1157D, 2012MNRAS.426.1121C, 2012JCAP...05..003B, Fleury}. The results of \FLUletter also suggested that cosmic shear observations could be plagued with non-negligible finite-beam corrections. This was further investigated in a companion paper~\cite{shear}, hereafter \FLUshear; it turns out that finite-beam corrections were overestimated in \FLUletter, due to simplistic assumptions on the distribution of matter in the Universe.

Unlike infinitesimal sources, extended sources can exhibit more complex distortions than a mere shear. The notion of flexion~\cite{Goldberg:2004hh, Bacon:2005qr}, for example, which characterizes the arckiness of an image, has already been thoroughly investigated in the literature. In the present article, we propose a simple unified mathematical description of weak lensing beyond shear, thereby generalizing the theory of flexion, and apply it to a realistic cosmological model. Furthermore, while the analysis of \FLUletter was limited to circular sources only, we show how our finite-beam formalism can be generalized to noncircular sources.

The article is organized as follows: in Sec.~\ref{sec:formalism}, we summarize the general context, approximations, and equations of the finite-beam formalism; in Sec.~\ref{sec:moments}, we show how the notion of shear can be extended to higher-order moments of an image; we compute these higher moments in a cosmological context in Sec.~\ref{sec:cosmic}, as well as their two-point correlations, in the case of circular sources; finally, we show how to tackle noncircular sources in Sec.~\ref{sec:noncircular}, and conclude in Sec.~\ref{sec:conclusion}.

We adopt units in which the speed of light is unity. Two-dimensional vectors are denoted with bold symbols ($\vect{\beta}, \vect{\theta}, \vect{\lambda}, \ldots$) while underlined quantities ($\cplx{\beta}, \cplx{\theta}, \cplx{\lambda}, \ldots$) are their complex representation: if $\vect{\theta}=(\theta_x, \theta_y)$, then $\cplx{\theta}\define\theta_x + \i \theta_y$.

\section{Formalism}
\label{sec:formalism}

This section briefly exposes our finite-beam formalism; further details about its construction, including physical motivations, can be found in \FLUshear~\cite{shear}. We consider a statistically homogeneous and isotropic Universe, filled with noncompact, spherical, nonrotating, and slowly moving massive objects (apart from their cosmic recession). The geometry of the resulting spacetime can be modeled by the Friedmann-Lema\^itre-Robertson-Walker (FLRW) metric with scalar perturbations,
\begin{multline}\label{eq:metric}
\dd s^2 = a^2(\eta) \Big\{ -(1+2\Phi)\dd\eta^2 \\
											+ (1-2\Phi) \pac{\dd\chi^2 + f_K^2(\chi) \, \dd\Omega^2 } \Big\} \ ,
\end{multline}
where $a$ denotes the scale factor quantifying cosmic expansion, $K$ is the background spatial curvature parameter, $f_K(\chi) \define \sin(\sqrt{K}\chi)/\sqrt{K}$, $\Phi$ is the gravitational potential generated by the massive objects, and $\eta, \chi$ are respectively the background conformal time and comoving radial coordinate.

\begin{figure}[h!]
\centering
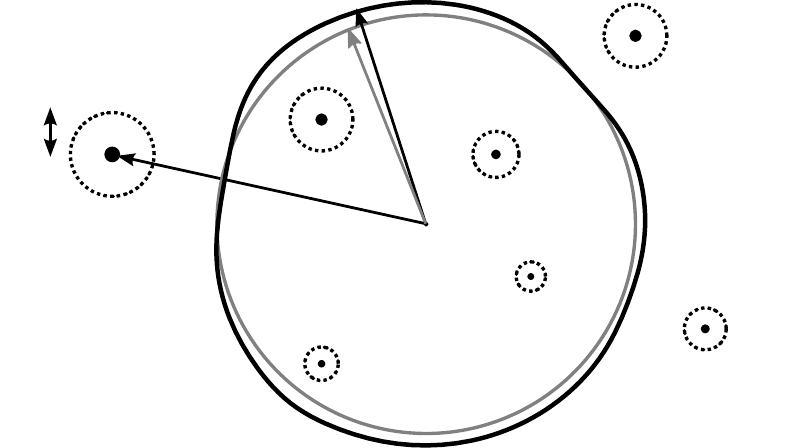\\[1cm]
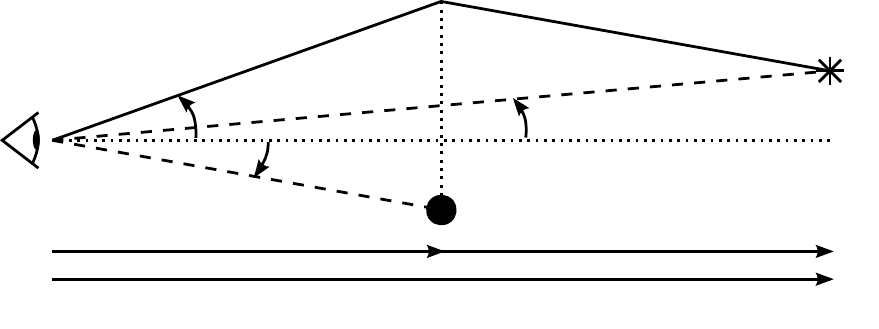
\caption{Top panel: Image $\image$ of an extended source~$\source$ by multiple weak lenses with Einstein radii; $\vect{\theta}$ denotes the main image of a point-source~$\vect{\beta}$, and $\eps_k, \vect{\lens}_k$ respectively denote the Einstein radius and angular position of the lens $k$. Bottom panel: Geometrical quantities involved in the lens equation.}
\label{fig:extended_source}
\end{figure}

Let an extended source be made of points which, in the absence of lensing, i.e. in a strictly homogeneous Universe, are observed in directions $\vect{\beta}$, as depicted in Fig.~\ref{fig:extended_source}. Although $\vect{\beta}$ represents an angular difference between two positions on the observer's celestial sphere, we will assume that this angle is small enough for $\vect{\beta}$ to be well approximated by a vector in a plane. In other words, the source is extended, but small, so that paraxial optics (flat-sky approximation) is valid. Let us call~$\source$ the unlensed contour of the source. If point-lenses are placed at various positions~$\vect{\lens}_k$, then the image~$\vect{\theta}$ of a point-source at $\vect{\beta}$ satisfies the \emph{lens equation}
\begin{equation}\label{eq:lens_equation}
\vect{\beta} = \vect{\theta} 
- \sum_{k} \eps_k^2 \, \frac{\vect{\theta}-\vect{\lens}_k}{\abs{\vect\theta-\vect{\lens}_k}^2} \ ,
\end{equation}
where $\eps_k$ denotes the \emph{Einstein radius} of the lens $k$. The Einstein radius of a lens quantifies its capacity to distort images.

Equation~\eqref{eq:lens_equation} generally has several solutions: a given point source can be multiply imaged. In this article, we will restrict to the \emph{weak-lensing regime}, and only consider the main image~$\vect{\theta}$ of each point~$\vect{\beta}$. This regime is equivalent to considering that the distance between a point source and any lens is much larger than the Einstein radius of the lens, $|\vect{\beta}-\vect{\lambda}_k|\gg\eps_k$. The source-image displacement is then very small, and can be approximated as
\begin{equation}\label{eq:lensing_displacement}
\delta\vect{\theta}
\define \vect{\theta} - \vect{\beta} 
\approx \sum_k \eps_k^2 \, \frac{\vect{\beta}-\vect{\lambda}_k}{|\vect{\beta}-\vect{\lambda}_k|^2} \ .
\end{equation}
Because we only consider the main image of each point, the contour~$\source$ of the extended source is lensed into a slightly distorted contour~$\image$, as shown in Fig.~\ref{fig:extended_source}.

The Einstein radius of the lens~$k$ reads
\begin{equation}\label{eq:Einstein_radius}
\eps_k^2 \define \frac{4G m_k D_{k\mathrm{S}}}{D_{\mathrm{O}k} D\e{OS}}
= \frac{4 G m_k (1+z_k) f_K(\chi\e{S}-\chi_k)}{f_K(\chi\e{S}) f_K(\chi_k)} \ ,
\end{equation}
where $m_k$ is the mass of the lens, while $D_{k\mathrm{S}}$, $D_{\mathrm{O}k}$, and $D\e{OS}$ are the angular-diameter distances, respectively, of the source seen from the lens $k$, of the lens~$k$ seen from the observer, and of the source seen from the observer; $z_k$ denotes the observed redshift of the lens.

Quite importantly, because we have chosen the background spacetime to be FLRW, which corresponds to a Universe homogeneously filled with matter, the mass of the lenses~$m_k$ and the corresponding squared Einstein radius, $\eps_k^2$, are allowed to be \emph{negative}. This is due to the fact that $\Phi$, which drives light deflection with respect to this background, satisfies a Poisson equation of the form~$\Delta\Phi = 4\pi G a^{2}(\rho-\bar{\rho})$, where $\bar{\rho}$ is the mean energy density. Introducing negative masses is a trick to account for the presence of $-\bar{\rho}$ in this equation; see Appendix~A of \FLUshear. For that reason, when going from a discrete to a continuous description of the matter distribution in Sec.~\ref{subsec:discrete_to_continuous}, the masses $m_k$ of the lenses will be replaced by $\delta\rho\define\rho-\bar{\rho}$, instead of simply $\rho$.

It is convenient to adopt a complex representation $\cplx{\beta}, \cplx{\theta}, \cplx{\lens}_k$ of the two-dimensional vectors~$\vect{\beta}, \vect{\theta}, \vect{\lens}_k$. If $(\vect{e}_x, \vect{e}_y)$ denotes an arbitrary orthonormal basis of the flat sky, then
\begin{equation}
\vect{\theta} = \theta_x \vect{e}_x + \theta_y \vect{e}_y
\longmapsto
\cplx{\theta} = \theta_x + \i \theta_y \ .
\end{equation}
With this notation, the lensing displacement~\eqref{eq:lensing_displacement} becomes~\cite{1973ApJ...185..747B}
\begin{equation}\label{eq:lens_equation_complex}
\delta\cplx{\theta}
= \cplx{\theta} - \cplx{\beta}
= \sum_k \frac{\eps_k^2}{\cplx{\beta}^* - \cplx{\lens}_k^*} \ ,
\end{equation}
where a star denotes complex conjugation.

The complex notation is particularly useful for describing the distortions of an image. Its most straightforward application is the calculation of the convergence~$\kappa=(\Omega-\Omega\e{S})/(2\Omega\e{S})$, where $\Omega, \Omega\e{S}$ respectively denote the angular area of the image and the source. From
\begin{equation}\label{eq:Omega_def}
\Omega
= \int_{\interior\image} \dd^2 \vect{\theta}
= \frac{1}{2\i} \ointctrclockwise_{\mathcal{I}} \cplx{\theta}^* \dd \cplx{\theta} \ ,
\end{equation}
one can substitute the complex lens equation, apply the residue theorem, and find~[FLU17]
\begin{equation}\label{eq:convergence}
\kappa = \sum_{k\in\interior\mathcal{S}} \frac{\pi \eps_k^2}{\Omega\e{S}} \ ,
\end{equation}
at lowest order in the lensing displacement $\delta\vect{\theta}$. This result shows in particular that, at this order of approximation, only the lenses enclosed by the light beam---interior lenses---contribute to its focusing. The case of shear is comprehensively investigated in \FLUshear, showing the respective roles of interior and exterior lenses. In the present article, we generalize the analysis of \FLUshear by showing how the complex formalism allows one to elegantly calculate all the moments of an image, thereby characterizing their shape with precision.

\section{Moments of an image}
\label{sec:moments}

Measurements of the weak gravitational shear are historically based on the image quadrupole (or second moment),
\begin{equation}\label{eq:quadrupole_def}
\quadrupole_{ab} 
= \frac{\int W[I(\vect{\theta})] \, \theta_a \theta_b \; \dd^2\vect{\theta}}
			{\int W[I(\vect{\theta})] \; \dd^2\vect{\theta}} \ ,
\end{equation}
where $I(\vect{\theta})$ is the image surface brightness in the direction $\vect{\theta}$, and $W$ is a weighting function. The quadrupole matrix~$\mat{\quadrupole}$ is then used to define the ellipticity\footnote{The usual notation for this ellipticity is $\chi$, we chose to call it $E$ in order to avoid confusions with the comoving radial coordinate.} of the image as~\cite{1995A&A...294..411S}
\begin{equation}\label{eq:ellipticity_def}
E
\define \frac{2(\quadrupole_{\langle 11\rangle} + \i \quadrupole_{\langle 12\rangle})}
						{\tr \mat{\quadrupole}} \ ,
\end{equation}
where angular brackets $\quadrupole_{\langle ab \rangle}$ denotes the traceless part of $\mat{\quadrupole}$.
In this section, we propose a generalization of $\mathcal{\quadrupole}$ and $E$, in order to characterize distortions of the shape of extended sources beyond shear.

\subsection{Generalizing the image quadrupole}

For any strictly positive integer $n$, we define the \emph{image moments} as
\begin{equation}\label{eq:moment_def}
\moment_{a_1 \ldots a_n} \define
\frac{\int W[I(\vect{\theta})] \, \theta_{a_1} \ldots \theta_{a_n} \; \dd^2\vect{\theta}}
			{\int W[I(\vect{\theta})] \; \dd^2\vect{\theta}} \ .
\end{equation}
While the second moment (quadrupole) characterizes the ellipticity of the image, the third one (octupole) quantifies its triangularity, the fourth one (hexadecapole) its squarity, and so on. In Ref.~\cite{2007ApJ...660..995O}, moments beyond the quadrupole were dubbed higher-order lensing image's characteristics (HOLICs). In Eq.~\eqref{eq:moment_def}, we have set the origin of image positions~$\vect{\theta}$ at the $W$-center of the image, that is
\begin{equation}
\int W[I(\vect{\theta})] \, \vect{\theta} \; \dd^2\vect{\theta} = \vect{0} \ ,
\end{equation}
which implies that the first moment (dipole) $\moment_a$ is zero.

Following \FLUletter, we assume for simplicity that $W$ is a top-hat function with an arbitrary brightness threshold, so that $W=1$ within the image, and $W=0$ otherwise. The denominator of Eq.~\eqref{eq:moment_def} then becomes the angular area~$\Omega$ of the image, while the numerator can be turned into a one-dimensional integral over the contour~$\image$ of the image:
\begin{align}
\moment_{a_1 \ldots a_n}
&= \frac{1}{\Omega} \int_{\interior\image} \theta_{a_1} \ldots \theta_{a_n} \; \dd^2\vect{\theta} \\
&= \frac{1}{(n+2)\Omega} \int_{\interior\image}
		\pd{(\theta_{a_1} \ldots \theta_{a_n} \theta_b)}{\theta_b} \; \dd^2\vect{\theta} \label{eq:moment_calculus_1}\\
&= \frac{1}{(n+2)\Omega} \ointctrclockwise_{\image}
		\theta_{a_1} \ldots \theta_{a_n} \; \det(\vect{\theta}, \dd\vect{\theta}) \label{eq:moment_calculus_2}\\
&= \frac{1}{(n+2)\Omega} \int_0^{2\pi}
		\hat{\theta}_{a_1}\ldots \hat{\theta}_{a_n} \, \theta^{n+2}\; \dd\psi \ ,
\end{align}
where we used Stokes' theorem to go from Eq.~\eqref{eq:moment_calculus_1} to Eq.~\eqref{eq:moment_calculus_2}, and in the last line we introduced the norm~$\theta$ of $\vect{\theta}$, and the unit vector~$\hat{\vect{\theta}}\define \vect{\theta}/\theta$ with components $\hat{\vect{\theta}}=(\cos\psi, \sin\psi)$.

The fully symmetric tensor~$\mat{\moment}$ has, in general, $n+1$ independent components, but this number drops to $2$ if we only consider its trace-free part, $\moment_{\langle a_1 \ldots a_n \rangle}$, whose contraction of any pair of indices vanishes,
\begin{equation}\label{eq:trace_free}
\forall i\not=j \qquad \delta^{a_i a_j} \moment_{\langle a_1 \ldots a_n \rangle} = 0 \ .
\end{equation}
Since the left-hand side of Eq.~\eqref{eq:trace_free} is a symmetric tensor with $n-2$ indices, the above represents $n-1$ independent constraints, whence the fact that $\moment_{\langle a_1 \ldots a_n \rangle}$ has only two independent components. These can be chosen as $\moment_{\langle 1 \ldots 11 \rangle}$ and $\moment_{\langle 1 \ldots 12 \rangle}$. Indeed, any other component will have pairs of indices with the value $2$, which can thus be converted into pairs of $1$ by the trace-free condition, and reshuffled in order to get either of the two aforementioned components. From these two independent components, we define the \emph{complex moment}
\begin{equation}
M_n \define \moment_{\langle 1 \ldots 11 \rangle} + \i \moment_{\langle 1 \ldots 12 \rangle} \ ,
\end{equation}
which is a direct generalization of the numerator of the complex ellipticity~\eqref{eq:ellipticity_def} of an image.

The final step consists in using that
\begin{align}
\hat{\theta}_{\langle 1} \ldots \hat{\theta}_{1}\hat{\theta}_{1\rangle} 
=& \frac{\cos n\psi}{2^{n-1}} \ ,\\
\hat{\theta}_{\langle 1} \ldots \hat{\theta}_{1}\hat{\theta}_{2\rangle} 
=& \frac{\sin n\psi}{2^{n-1}} \ ,
\end{align}
where it is understood that the left-hand sides contain $n$ factors. These relations are relatively well known in the framework of symmetric-trace free tensors; they can be proved by induction. The complex representation~$\cplx{\theta}$ of $\vect{\theta}$ then naturally arises into the expression of $M_n$,
\begin{align}
M_n
&= \frac{1}{2^{n-1}(n+2)\Omega} \int_0^{2\pi} \theta^{n+2} \ex{\i n\psi} \; \dd\psi \\
&= \frac{1}{2^{n-1}(n+2)\Omega} \int_0^{2\pi} \theta^2 \cplx{\theta}^n \; \dd\psi \ .
\end{align}

Why only consider trace-free moments? In fact, this is only justified in the case of circular sources. Consider a circular source with constant radius~$\beta$, and write $\vect{\theta}=\vect{\beta}+\delta\vect{\theta}$, for each point~$\vect{\beta}=\beta\,\hat{\vect{\beta}}$ of this circle. Then the $n$th moment reads
\begin{multline}\label{eq:moment_trace}
\moment_{a_1 \ldots a_n}
= \frac{\beta^{n+2}}{(n+2)\Omega} \int_0^{2\pi} 
															\hat{\theta}_{a_1} \ldots \hat{\theta}_{a_n} \; \dd\psi \\
	+ \frac{\beta^{n+1}}{2\Omega} \int_0^{2\pi} 
		\pa{ \hat{\vect{\beta}}\cdot \delta\vect{\theta} } \hat{\theta}_{a_1} \ldots \hat{\theta}_{a_n} \; \dd\psi 
	+ \mathcal{O}(\delta\theta^2) \ .
\end{multline}
Since $\delta^{ab}\hat{\theta}_{a}\hat{\theta}_{b}=1$, any trace of the $n$th moment $\moment_{a_1 \ldots a_n}$ is related to 
the $(n-2)$th moment; hence our interest in the trace-free part. This rationale, however, does not hold if the source is not circular, and thus we lose information by focusing on the trace-free moments in general.

The complex ellipticity~\eqref{eq:ellipticity_def} is a normalized version of $M_2$, using $\tr\mat{\quadrupole}$ to eliminate the direct dependency in the area of the image. Similarly, we choose to normalize $M_n$ with
\begin{equation}
\normalization_n
\define \frac{n+1}{2^n \Omega} \int_{\interior\image} \theta^n \; \dd^2\vect{\theta}
= \frac{1}{2^n \Omega} \int_0^{2\pi} \theta^{n+2} \; \dd\psi \ ,
\end{equation}
thereby defining the \emph{reduced $n$th moment} of the image,
\begin{empheq}[box=\fbox]{equation}\label{eq:reduced_moment}
\mu_n
\define \frac{M_n}{\normalization_n}
=  \frac{2}{n+2}\,\frac{\int_0^{2\pi} \cplx{\theta}^n \; \theta^2 \dd\psi}
										{\int_0^{2\pi} \theta^n \; \theta^2\dd\psi} \ .
\end{empheq}
Recall that $n>0$, and that by construction $\mu_1=0$. The first reduced moment containing information is thus $\mu_2$, which corresponds to the complex ellipticity, $E=2\mu_2$. As will be further discussed in Sec.~\ref{subsec:comparison_flexion_roulettes}, $\mu_3$ is related to the so-called $\mathcal{G}$-type flexion~\cite{Bacon:2005qr}. To our knowledge, the moments $n\geq 4$ have never been considered in the weak-lensing literature.

\subsection{Expression of the reduced moments in weak lensing}
\label{subsec:reduced_moments_weak_lensing}

Let us now relate the reduced moments~$\mu_n$ to the properties of the source~$\source$ and of the lenses which turn it into the image~$\image$. Our goal is to derive an expression of the form~$\mu_n = \mu_n\h{S} + \delta\mu_n$, where $\mu_n\h{S}$ is the intrinsic reduced moment of the source, and $\delta\mu_n$ its observed correction due to lensing. We start with the complex expression of the lens equation
\begin{equation}
\cplx{\theta} = \cplx{\beta} + \delta\cplx{\theta} \ ,
\end{equation}
and expand the integrals of Eq.~\eqref{eq:reduced_moment} at first order in $\delta\cplx{\theta}$, starting with the numerator. On the one hand, the integrand reads
\begin{align}
\cplx{\theta}^n\theta^2
= \cplx{\beta}^n\beta^2 +
		\cplx{\beta}^n \pac{(n+1)\cplx{\beta}^*\delta\cplx{\theta} + \cplx{\beta}\delta\cplx{\theta}^* } .
\end{align}
On the other hand, we must be careful of the fact that integration is performed over the polar angle $\psi$ of the image points~$\cplx{\theta}=\theta\ex{\i\psi}$, which differs from the polar angle~$\ph$ of the corresponding source points~$\cplx{\beta}=\beta\ex{\i\ph}$. Defining $\delta\psi \define \psi-\ph$, and writing that, at first order in $\delta\psi$, $\cplx{\theta}=\theta\ex{\i\ph}(1+\i\delta\psi)=\cplx{\beta}+\delta\cplx{\theta}$, we find
\begin{equation}
\delta\psi = \Im\pa{\cplx{\beta}^{-1}\delta\cplx{\theta}} .
\end{equation}
The integral of $M_n$ thus reads, at first order,
\begin{multline}
\int_0^{2\pi} \cplx{\theta}^n \theta^2 \, \dd\psi
= \int_0^{2\pi} \cplx{\beta}^n \beta^2 \pa{1+\ddf{\delta\psi}{\ph}}\dd\ph  \\
	+ \int_0^{2\pi} \cplx{\beta}^n \pac{(n+1)\cplx{\beta}^*\delta\cplx{\theta} + \cplx{\beta}\delta\cplx{\theta}^* } \dd\ph \ .
\end{multline}
Integrating the first term by parts, replacing $\delta\psi$ with its expression, and rearranging the various terms, we get
\begin{multline}\label{eq:reduced_moment_calculation}
\int_0^{2\pi} \cplx{\theta}^n \theta^2 \, \dd\psi
= \int_0^{2\pi} \cplx{\beta}^n \beta^2 \, \dd\ph \\
	+ \frac{n+2}{2\i} \int_0^{2\pi} \delta\cplx{\theta}^* \, \cplx{\beta}^n \pa{\ddf{\beta}{\ph}+\i\beta} \ex{\i\ph} \; \dd\ph \\
	- \frac{n+2}{2\i} \int_0^{2\pi} \delta\cplx{\theta} \, \cplx{\beta}^n \pa{\ddf{\beta}{\ph}-\i\beta} \ex{-\i\ph} \; \dd\ph \ .
\end{multline}
The final step consists in recognizing, in the last two terms of Eq.~\eqref{eq:reduced_moment_calculation}, the differential $\dd\cplx{\beta}=(\dd\beta/\dd\ph + \i\beta)\ex{i\ph}\dd\ph$ and its complex conjugate. In other words, we have
\begin{multline}\label{eq:reduced_moment_numerator_result}
\int_0^{2\pi} \cplx{\theta}^n \theta^2 \, \dd\psi
= \int_0^{2\pi} \cplx{\beta}^n \beta^2 \, \dd\ph \\
	+ \frac{n+2}{2\i} \ointctrclockwise_{\source} \delta\cplx{\theta}^* \cplx{\beta}^n \dd\cplx{\beta}
	+ \pac{\frac{n+2}{2\i} \ointctrclockwise_{\source} \delta\cplx{\theta}^* (\cplx{\beta}^*)^n \dd\cplx{\beta}}^* \ .
\end{multline}

The calculation of the denominator of Eq.~\eqref{eq:reduced_moment}, corresponding to the normalization~$\normalization_n$, follows similar lines, and yields
\begin{equation}\label{eq:reduced_moment_denominator_result}
\int_0^{2\pi} \theta^{n+2} \, \dd\psi
= \int_0^{2\pi} \beta^{n+2} \, \dd\ph
	+ 2\,\Re \pa{ \frac{n+2}{2\i} \ointctrclockwise_{\source} \delta\cplx{\theta}^* \beta^n \dd\cplx{\beta} } \ .
\end{equation}
Gathering Eqs.~\eqref{eq:reduced_moment_numerator_result} and \eqref{eq:reduced_moment_denominator_result}, we obtain
\begin{multline}\label{eq:reduced_moment_result}
\mu_n
= \pac{
			1 - \frac{(n+2)\Re\pa{ \frac{1}{2\pi\i} \ointctrclockwise_{\source} \delta\cplx{\theta}^* \beta^n \dd\cplx{\beta} }}
							{\frac{1}{2\pi} \int_0^{2\pi} \beta^{n+2} \, \dd\ph} 
			} \mu_n\h{S} \\
		+ \frac{ \frac{1}{2\pi\i} \ointctrclockwise_\source \delta\cplx{\theta}^* \cplx{\beta}^n \, \dd\cplx{\beta} 
					 + \pac{\frac{1}{2\pi\i} \ointctrclockwise_\source \delta\cplx{\theta}^* (\cplx{\beta}^*)^n \, \dd\cplx{\beta} }^* }
					{ \frac{1}{2\pi} \int_0^{2\pi} \beta^{n+2} \dd\ph } \ ,
\end{multline}
which shows shows how weak lensing affects the reduced multipole of an image at lowest order in light deflection. The advantage of this expression is that all the lensing effects are expressed in terms of complex integrals of $\delta\cplx{\theta}^*$. By virtue of the lens equation, this quantity reads, still at lowest order,
\begin{equation}
\delta\cplx{\theta}^* = \sum_{k=1}^{\infty} \frac{\eps_k^2}{\cplx{\beta}-\cplx{\lambda}_k} \ .
\end{equation}
Therefore, $\mu_n$ takes the form

\vspace*{2mm}

\noindent
\fbox{
\begin{minipage}{0.95\columnwidth}
\vspace*{-4mm}
\begin{multline}\label{eq:reduced_moment_ABCD}
\mu_n
= \paac{
			1 - \frac{n+2}{D_n} \, \sum_k \eps_k^2 \Re\pac{C_n(\cplx{\lambda}_k)}
			} \mu_n\h{S} \\
		+ \frac{1}{D_n} \paac{ \sum_k \eps_k^2 A_n(\cplx{\lambda}_k) 
											+ \sum_k \eps_k^2 \pac{B_n(\cplx{\lambda}_k)}^* } ,
\end{multline}
with the four integrals
\begin{align}
\label{eq:A_n}
A_n(\cplx{\lambda})
&\define \frac{1}{2\pi \i}
\ointctrclockwise_\source \frac{\cplx{\beta}^n \, \dd\cplx{\beta}}{\cplx{\beta}-\cplx{\lambda}} \ , \\
\label{eq:B_n}
B_n(\cplx{\lambda})
&\define \frac{1}{2\pi \i}
\ointctrclockwise_\source \frac{(\cplx{\beta}^*)^n \, \dd\cplx{\beta}}{\cplx{\beta}-\cplx{\lambda}} \ , \\
\label{eq:C_n}
C_n(\cplx{\lambda})
&\define \frac{1}{2\pi \i}
\ointctrclockwise_\source \frac{\beta^n \, \dd\cplx{\beta}}{\cplx{\beta}-\cplx{\lambda}} \ , \\
\label{eq:D_n}
D_n &\define \frac{1}{2\pi} \int_0^{2\pi} \beta^{n+2} \, \dd\ph \ .
\end{align}
\end{minipage}
}

\vspace*{2mm}

Determining $\mu_n$ for a given source~$\source$ thus consists in computing the integrals~$A_n, B_n, C_n, D_n$.
While $A_n$ is directly integrated via the residue theorem, the last two are more challenging in general. We will see in Sec.~\ref{sec:noncircular} how to handle them, and focus on circular sources for the remainder of this section.

\subsection{Circular sources}
\label{subsec:circular_sources}

The results which have been obtained so far are fully general with respect to the shape~$\source$ of the source. However, they greatly simplify, and are more easily interpreted, in the case of circular sources. Thus, \emph{from now on and until Sec.~\ref{sec:noncircular}, we restrict the analysis to circular sources}, i.e. $\cplx{\beta}=\beta\ex{\i\ph}$ where $\beta$ does not depend on $\ph$. It is easy to see that all the intrinsic moments of a circular source vanish, $\mu_n\h{S}=0$ for any $n>1$. The moments of the image are then all due to lensing, and read
\begin{equation}\label{eq:moment_circular_source}
\mu_n
= \frac{1}{\beta^{n+2}}
	\sum_k \eps_k^2 \paac{ A_n(\cplx{\lambda}_k) + \pac{ B_n(\cplx{\lambda}_k) }^* } \ .
\end{equation}
The residue theorem immediately yields, for the first integral,
\begin{equation}\label{eq:moment_circular_source_first_integral}
A_n(\cplx{\lambda}_k)
\define
\frac{1}{2\pi\i}
\ointctrclockwise_\source \frac{\cplx{\beta}^n \, \dd\cplx{\beta}}{\cplx{\beta}-\cplx{\lambda}_k}
=
\begin{cases}
\cplx{\lambda}_k^n & \text{if } \cplx{\lambda}_k \in \interior\source \\
0 & \text{if } \cplx{\lambda}_k \in \exterior\source \ ,
\end{cases}
\end{equation}
so that only the lenses enclosed by the source contribute to this part of $\mu_n$ (this also holds for noncircular sources).

The $B_n$ integral is less immediately calculated, because its integrand is not obviously $\mathbb{C}$-differentiable: it depends on both $\cplx{\beta}$ and $\cplx{\beta}^*$. However, since $\source$ is a circle, for any $\cplx{\beta}\in\source$ we can write $\cplx{\beta}^*=\beta^2/\cplx{\beta}$, where $\beta^2$ is a constant which can be taken out of the integral,
\begin{equation}\label{eq:B_n_circular}
B_n(\cplx{\lambda}_k)
\define
\frac{1}{2\pi\i} \ointctrclockwise_\source \frac{(\cplx{\beta}^*)^n \, \dd\cplx{\beta}}{\cplx{\beta}-\cplx{\lambda}_k}
= \frac{\beta^{2n}}{2\pi\i} \ointctrclockwise_\source \frac{\dd\cplx{\beta}}{\cplx{\beta}^n(\cplx{\beta}-\cplx{\lambda}_k)} \ .
\end{equation}
The residue theorem can now be applied, either directly, allowing for the fact that the integrand generally has two poles (at $0$ and $\cplx{\lambda}_k$), or after changing the variable to $w=\beta/\cplx{\beta}$ which brings one back to Eq.~\eqref{eq:moment_circular_source_first_integral}. The result is
\begin{equation}
\frac{1}{2\pi\i}
\ointctrclockwise_\source \frac{\dd\cplx{\beta}}{\cplx{\beta}^n(\cplx{\beta}-\cplx{\lambda}_k)}
=
\begin{cases}
0 & \text{if } \cplx{\lambda}_k \in \interior\source \\
- \cplx{\lambda}_k^{-n}  & \text{if } \cplx{\lambda}_k \in \exterior\source \ ,\\
\end{cases}
\end{equation}
hence this second contribution only depends on the lenses located outside the source.

Summarizing, the reduced moments of the image of a weakly lensed circular source read
\begin{empheq}[box=\fbox]{equation}\label{eq:reduced_moment_circular_source_result}
\mu_n
=
\underbrace{
					\sum_{k\in\interior\source} \pa{\frac{\eps_k}{\beta}}^2 \pa{ \frac{\cplx{\lambda}_k}{\beta} }^n 
					}_{\mu_n\h{int}}
\underbrace{
					- \sum_{k\in\exterior\source} \pa{\frac{\eps_k}{\beta}}^2 \pa{ \frac{\beta}{\cplx{\lambda}_k^*} }^n 
					}_{\mu_n\h{ext}}
\ .
\end{empheq}
This generalizes the case of shear, $\gamma = \mu_2$, obtained in \FLUshear. The closer a lens is to the contour of the source, in angle space, the greater its impact on the image moments. This behavior is enhanced as $n$ is larger, so that large moments are only sourced by lenses which are very close to the source in angle space.

\subsection{Relation between reduced moments and Fourier modes}
\label{subsec:Fourier}

The geometric meaning of the reduced moments is clearer when reinterpreted as a combination of Fourier modes. Consider again a circular source, and let us parametrize the displacement~$\delta\cplx{\theta}=\cplx{\theta}-\cplx{\beta}$ of its contour with the polar angle~$\ph$ of the associated point source~$\cplx{\beta}=\beta\ex{\i\ph}$. Since $\source$ is a closed curve, the complex function~$\ph\mapsto\delta\cplx{\theta}(\ph)$ is $2\pi$-periodic, and thus it can be expanded in Fourier series as
\begin{align}
\delta\cplx{\theta}(\ph)
&= \sum_{p\in\mathbb{Z}} \delta\cplx{\theta}_p \, \ex{\i(p+1)\ph} \ ,\\
\delta\cplx{\theta}_p
&\define \frac{1}{2\pi} \int_0^{2\pi} \delta\cplx{\theta} \, \ex{-\i(p+1)\ph} \; \dd\ph \ .
\end{align}
Comparing the definition of $\delta\cplx{\theta}_p$ with the expression~\eqref{eq:reduced_moment_result}, one immediately sees that, for any $n>1$,
\begin{equation}\label{eq:reduced_moment_Fourier}
\mu\h{int}_n = \frac{\delta\cplx{\theta}_{n}^*}{\beta} \ ,
\quad \text{and} \quad
\mu\h{ext}_n = \frac{\delta\cplx{\theta}_{-n}}{\beta} \ .
\end{equation}
This shows that, as far as circular sources are concerned, interior lenses (i.e. lenses enclosed by the source) only generate positive Fourier modes of distortion, while exterior lenses only generate negative Fourier modes, as already noticed in \FLUletter. Note however that those modes are not individually observable, because the polar angle~$\ph$ itself is not observable. In other words, measuring a slightly triangular image shape ($\mu_3$) does not tell one whether it is due to an interior lens ($\mu_3\h{int}=\delta\cplx{\theta}_{3}^*/\beta$) or an exterior lens ($\mu_3\h{ext}=\delta\cplx{\theta}_{-3}/\beta$). The other moments can be used to break this degeneracy, because the dependence of $\mu_n\h{int}$ in the position of interior lenses is different from the dependence of $\mu_n\h{ext}$ in the position of exterior lenses.

\begin{figure*}[ht]
\centering
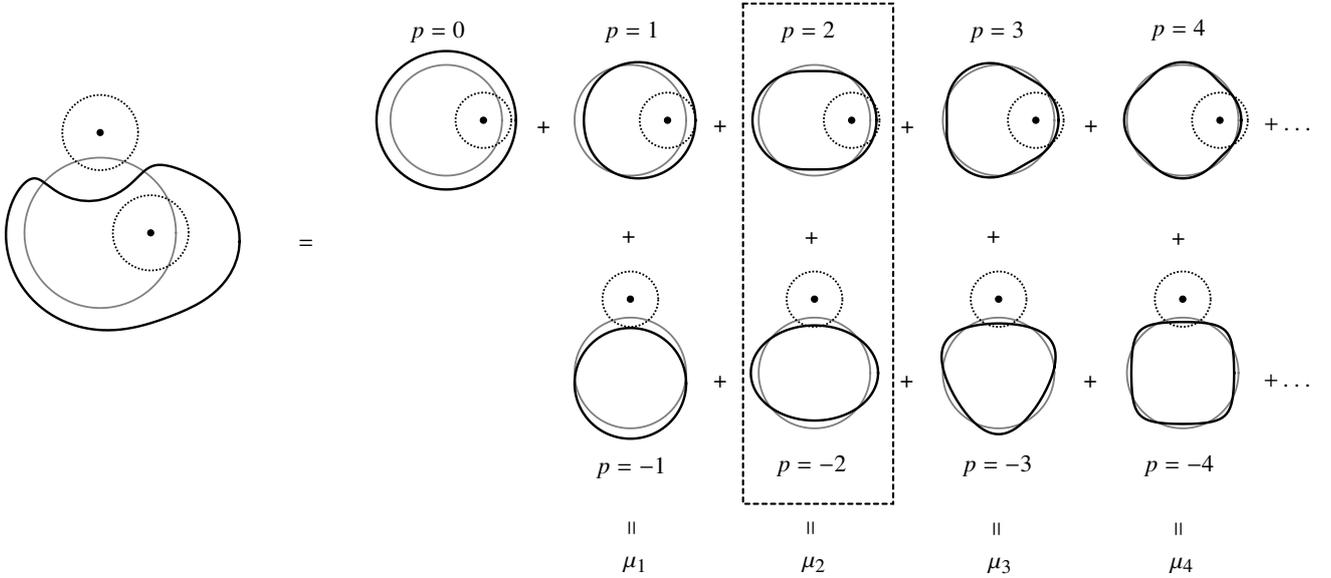
\caption{Fourier decomposition of the image (black line) of a circular source (gray line) by a couple of lenses (black dots), whose Einstein radii are indicated with dotted lines. We chose here to go far beyond the weak-lensing regime, $|\vect{\beta}-\vect{\lambda}|\e{min}=2\eps/3$, in order to visually enhance the effect. On the right-hand side, each black line shows the effect of a single Fourier mode~$p$, i.e. $\ph\mapsto\beta\ex{\i\ph}+\delta\cplx{\theta}_p \ex{\i(p+1)\ph}$. By virtue of Eq.~\eqref{eq:reduced_moment_Fourier}, for any $n>0$, the $n$th reduced complex moment is given by the combination of the Fourier modes $n$ and $-n$, $\mu_n=(\delta\cplx{\theta}_n^*+\delta\cplx{\theta}_{-n})/2$.}
\label{fig:Fourier_modes}
\end{figure*}

Despite the fact that the Fourier modes $\delta\cplx{\theta}_n$ are not individually observable, they are convenient for visualizing the respective effect of interior and exterior lenses on a circular source. Indeed, the contour of the image
\begin{equation}
\cplx{\theta}(\ph) = \beta\ex{\i\ph} + \sum_{p\in\mathbb{Z}} \delta\cplx{\theta}_p \ex{\i(p+1)\ph}
\end{equation}
can be viewed as a curve drawn by a fictitious device made of successive wheels with different sizes and spinning with different angular velocities. Suppose, for example, that there is only a single nonvanishing mode $\delta\cplx{\theta}_p$. Now consider a wheel with radius $\beta$; on the surface of this first wheel, fix the center of second wheel with radius $|\delta\cplx{\theta}_p|$, and on the surface of this second wheel, attach a pen at angular position $\Arg(\delta\cplx{\theta}_p)$. Then $\image$ is the curve drawn by the pen if the first wheel rotates with angular velocity~$\omega$, thereby dragging the center of the second wheel spinning with angular velocity~$(p+1)\times\omega$. The effect of the first four positive and negative modes is depicted in Fig.~\ref{fig:Fourier_modes}.

The relation~\eqref{eq:reduced_moment_Fourier} between reduced moments~$\mu_n$ and Fourier modes~$\delta\cplx{\theta}_p$ also provides an alternative way to compute~$\mu_n$ for circular sources. Indeed, the displacement field~$\delta\cplx{\theta}(\cplx{\beta})$ is nothing but a geometric series,
\begin{align}
\delta\cplx{\theta}
&= \sum_k \frac{\eps_k^2}{\cplx{\beta}^*-\cplx{\lambda}_k^*}\\
&= \sum_{k\in\interior\source} \frac{\eps_k^2}{\cplx{\beta}^*} \frac{1}{1 - \cplx{\lambda}_k^*/\cplx{\beta}^*}
	- \sum_{k\in\exterior\source} \frac{\eps_k^2}{\cplx{\lambda}^*_k} \frac{1}{1 - \cplx{\beta}^*/\cplx{\lambda}^*_k} \\
\nonumber
&= \sum_{p=0}^{\infty} 
		\pac{ \sum_{k\in\interior\source} \frac{\eps_k^2}{\beta} \pa{\frac{\cplx{\lambda}^*}{\beta}}^p }
		\ex{\i(p+1)\ph}\\
&\hspace*{2cm}
	-\sum_{p=-\infty}^{-1} 
		\pac{ \sum_{k\in\exterior\source} \frac{\eps_k^2}{\beta} \pa{\frac{\beta}{\cplx{\lambda}^*}}^p }
		\ex{\i(p+1)\ph} \ ,
\end{align}
where the Fourier modes~$\delta\cplx{\theta}_{p\geq0}$, $\delta\cplx{\theta}_{p<0}$, and hence $\mu_n\h{int}, \mu_n\h{ext}$,  can directly be read.

\subsection{Relation with flexion and Clarkson's roulettes}
\label{subsec:comparison_flexion_roulettes}

Weak lensing beyond shear has already been investigated in the literature, notably through the notion of \emph{flexion}. The first type of flexion, denoted by $\flexionF$, was first introduced by Goldberg and Bacon in Ref.~\cite{Goldberg:2004hh}, and the second type, denoted $\flexionG$, by Bacon \textit{et al.} in Ref.~\cite{Bacon:2005qr}. The $\flexionF$-type flexion is a spin-$1$ quantity, and is related to the displacement of the centroid of an image with respect to its contour. The $\flexionG$-type flexion has spin $3$, and can be seen as the triangularity, or arckiness, of the image. While the initial proposition for flexion measurements relied on shapelets~\cite{Refregier:2001fd, Refregier:2001fe}, another method, based on the image moments (HOLICs), was developed in Refs.~\cite{2007ApJ...660..995O, Okura:2007js, Okura:2008ch}, thereby extending a first analysis by other authors~\cite{Irwin:2005nc}. See Ref.~\cite{Goldberg:2006jp} for a comparison of the relative merits of shapelets and moments for flexion measurements.

In its standard formalism, flexion derives from shear. If $\gamma(\vect{\alpha})$ is the shear observed in a direction $\vect{\alpha}$, then\footnote{Differences with the original expressions of Ref.~\cite{Bacon:2005qr} come from (i) a different convention for shear; (ii) the fact that in this reference the complex derivative is defined in an unusual way,
\begin{equation*}
\partial \define \pd{}{\alpha_1} + \i \, \pd{}{\alpha_2} = 2 \, \pd{}{\cplx{\alpha}^*}
\end{equation*}
}
\begin{equation}
\flexionF \define -2 \, \pd{\gamma}{\cplx{\alpha}} \ , 
\qquad
\flexionG \define -2 \, \pd{\gamma}{\cplx{\alpha}^*} \ .
\end{equation}
These quantities are easily computed with our formalism: reintroducing the dependence in the observation direction~$\vect{\alpha}$ (center of the image) in Eq.~\eqref{eq:reduced_moment_circular_source_result}, and applying it to $n=2$ (shear), we indeed have
\begin{equation}
\gamma(\cplx{\alpha})
= \sum_{k\in\interior\source} \frac{\eps_k^2 (\cplx{\lambda}_k - \cplx{\alpha})^2}{\beta^4}
	- \sum_{k\in\exterior\source} \frac{\eps_k^2}{(\cplx{\lambda}_k^* - \cplx{\alpha}^*)^2} \ ,
\end{equation}
and hence
\begin{align}
\flexionF
&= \sum_{k\in\interior\source} \frac{4\eps_k^2 (\cplx{\lambda}_k - \cplx{\alpha})}{\beta^4}
=\frac{4\mu_1\h{int}}{\beta} \ ,\\
\flexionG
&= \sum_{k\in\exterior\source} \frac{4\eps_k^2}{(\cplx{\lambda}_k^* - \cplx{\alpha}^*)^3}
= -\frac{4\mu_3\h{ext}}{\beta} \ .
\end{align}
We thus recover the spin-$1$ and spin-$3$ properties of the two flexions, as well as their geometrical interpretation (see Fig.~\ref{fig:Fourier_modes}). The $\flexionF$-type flexion being only due to interior lenses, we recover the known fact that, outside of any form of matter, $\flexionF=0$. Since $\mu_1=0$ by definition, we conclude that the $\flexionF$-type flexion is not observable in our framework. This apparent contradiction with the literature, notably Refs.~\cite{2007ApJ...660..995O, Goldberg:2006jp}, is due to our restriction to a top-hat weighting function when calculating the moments~$M_n$, and hence $\mu_n$. This choice allowed us to turn the two-dimensional problem of the image analysis to a one-dimensional problem: the analysis of its contour. Albeit mathematically convenient, this restriction removes a part of the information contained in the image, notably the position of its centroid, which is precisely what $\flexionF$ acts on. Furthermore, in our framework, $\flexionG\propto\mu_3\h{ext}$ cannot be observed independently from its complementary term~$\mu_3\h{int}$, just like shear picks up contributions from interior lenses. We stress that, by construction, the standard flexion theory cannot allow for $\mu_3\h{int}$, which thus represents an entirely new component.

Let us close this section by discussing the connections between our approach and the recent work of Clarkson~\cite{2016CQGra..33pLT01C, 2016CQGra..33x5003C}---the so-called \emph{roulettes}. The roulette formalism somehow takes a path which is opposite to ours: while we use the strong-lensing formalism to describe weak lensing beyond infinitesimal beams (see Sec.~\ref{sec:formalism}), Clarkson extended the weak-lensing formalism to describe strong lensing. We thus expect both approaches to meet midway. In the roulette approach, the lensing displacement field~$\delta\vect{\theta}$ is computed via a nonlinear generalization of the geodesic deviation equation; the result takes the form (notations are adapted)
\begin{equation}\label{eq:decomposition_roulettes}
\delta\theta^a
= \sum_{m=1}^\infty \frac{1}{m!} \, \mathcal{A}\indices{^a_{b_1} _\ldots_{b_m}} \theta^{b_1} \ldots \theta^{b_m} \ ,
\end{equation}
where $\mathcal{A}\indices{^a_{b_1} _\ldots_{b_m}}$ is given by an integral of the transverse derivatives of the Riemann tensor. From the symmetric-trace-free part of $\mathcal{A}\indices{_a_{b_1} _\ldots_{b_m}}$, one then defines normal modes which appear to be very similar to the Fourier modes~$\delta\cplx{\theta}_p$ introduced in Sec.~\ref{subsec:Fourier} and depicted in Fig.~\ref{fig:Fourier_modes}. This similarity can be schematically explained as follows: in the weak-lensing regime, the $\theta$s in the right-hand side of Eq.~\eqref{eq:decomposition_roulettes} can be replaced by $\beta$s; then, modulo resummation, the traces of $\mathcal{A}\indices{^a_{b_1} _\ldots_{b_m}}$ can be absorbed in the terms $m-2$, $m-4$, etc. Calling $\mathcal{B}\indices{^a_\langle_{b_1} _\ldots_{b_m}_\rangle}$ the symmetric-trace-free tensors obtained after resummation,
\begin{align}
\delta\theta^a
&= \sum_{m=0}^\infty \frac{1}{m!} \, \mathcal{B}\indices{^a_\langle_{b_1} _\ldots_{b_m}_\rangle}
															 \hat{\beta}^{\langle b_1} \ldots \hat{\beta}^{b_m\rangle} \\
&= \sum_{m=0}^\infty \frac{1}{m!} \, 
		\mathcal{B}\indices{^a_\langle_1 _\ldots_1_1_\rangle} \cos m\ph
		+ \mathcal{B}\indices{^a_\langle_1 _\ldots_1_2_\rangle} \sin m\ph	\ .				 
\end{align}
Thus, there are combinations~$\hat{\mathcal{B}}_m$ of the components of the above tensors such that
\begin{equation}
\delta\cplx{\theta}
= \sum_{m\in\mathbb{Z}}^\infty \frac{\hat{\mathcal{B}}_m}{m!} \, \ex{\i m \ph} \ ,
\end{equation}
whence the correspondence between the Fourier modes~$\delta\cplx{\theta}_p$ and the normal modes of the roulette formalism.

\section{Cosmic weak lensing beyond shear}
\label{sec:cosmic}

Just like their apparent ellipticity, the other reduced moments~$\mu_n$ of images of galaxies are observable quantities, whose lensing contribution depends on the underlying distribution of matter in the Universe. This section generalizes what is currently the main observable of weak lensing---the shear two-point correlation function---to higher-order moments. For simplicity, the analysis is here restricted to circular sources, so that the results of Sec.~\ref{subsec:circular_sources} can be applied. Corrections due to noncircularity will be discussed in Sec.~\ref{sec:noncircular}.

\subsection{From discrete lenses to a continuous matter distribution}
\label{subsec:discrete_to_continuous}

In the previous section, we calculated the effect of a set of discrete lenses on an image's reduced moments. The first step towards cosmology consists in translating those results in terms of a continuous distribution of matter, described by a density field~$\rho$, rather than a list of masses and positions. This first step is identical to the cases of convergence and shear, and is extensively discussed in \FLUshear. The presentation will thus be slightly more laconic here.

Equation~\eqref{eq:reduced_moment_result} expresses the reduced moments~$\mu_n$ as sums of terms proportional to the lenses' squared Einstein radii~$\eps_k^2 \propto m_k$. Going from a discrete to a continuous model consists in turning sums into integrals, as
\begin{equation}
\sum_k m_k \, (\ldots) \rightarrow \int \dd^3 m \, (\ldots) = \int \delta\rho \, \dd^3 V \, (\ldots) \ ,
\end{equation}
where $\delta\rho \define \rho - \bar{\rho}$ is the density relative to the FLRW background. Remember that the masses~$m_k$ of the lenses were allowed to be negative, which explains why their continuous counterpart is $\delta\rho$ rather than $\rho$. See Appendix~A of \FLUshear for further details. Assuming that matter (which excludes dark energy) is nonrelativistic, we can write
$
\delta\rho \, \dd^3 V
= \bar{\rho}_0 \delta \, \dd^3 V_0
$, 
where $\delta$ denotes the density contrast, and a zero subscript indicates the value of a quantity today.

The philosophy of the continuous description is that, instead of summing over individual lenses with mass~$m_k$, comoving distance~$\chi_k$, and transverse (angular) position $\vect{\lambda}_k$, we sum over positions $\chi, \vect{\lambda}$ and count the mass comprised in an infinitesimal domain~$\dd^3 V_0$ about it. Introducing the polar angle~$\phi$ such that $\vect{\lambda}=\lambda(\cos\phi, \sin\phi)$, the volume element reads
\begin{equation}
\dd^3 V_0 = \dd\chi \times f_K(\chi) \dd\lens \times f_K(\chi)\lens \dd\phi
\end{equation}
in the flat-sky approximation ($\sin\lambda\approx\lambda$). Therefore, if $\vect{\alpha}$ denotes the direction of the center of the source, we have
\begin{multline}\label{eq:moment_fixed_z}
\mu_n(\chi\e{S},\vect{\alpha})
= 4\pi G \bar{\rho}_0 
	\int_0^{\chi\e{S}} \dd\chi \; (1+z) \, \frac{f_K(\chi\e{S}-\chi)f_K(\chi)}{f_K(\chi\e{S}} \\
	\times (\mathcal{K}_n * \delta)(\eta_0-\chi, \chi, \vect{\alpha}) \ ,
\end{multline}
which involves in the second line the convolution product
\begin{equation}
(\mathcal{K}_n * \delta)(\eta, \chi, \vect{\alpha})
\define \int_{\mathbb{R}^2} \frac{\dd^2\vect{\lambda}}{\Omega\e{S}} \; 
			\kernel_n(\vect{\lambda}) \delta(\eta, \chi, \vect{\alpha}+\vect{\lambda}) \ ,
\end{equation}
with the kernel~$\kernel_n = \kernel_n\h{int}+\kernel_n\h{ext}$,
\begin{align}
\kernel_n\h{int}(\vect{\lambda}) &= \Theta(\beta-\lambda) \pa{ \frac{\lambda}{\beta} }^n \ex{\i n \phi} \ , \\
\kernel_n\h{ext}(\vect{\lambda}) &= - \Theta(\lambda-\beta) \pa{ \frac{\beta}{\lambda} }^n \ex{\i n \phi} \ ,
\end{align}
where $\Theta$ is the Heaviside function. Since $\vect{\lambda}$ spans the positions of the lenses, $\Theta(\beta-\lambda)$ selects matter enclosed by the light beam, while $\Theta(\lambda-\beta)$ selects exterior matter.

\subsection{Effective moments}

When many sources are observed in the direction~$\vect{\alpha}$, it is customary to calculate their average moment in order to get rid of the dependence in $\chi\e{S}$. If $p(\beta, \chi_*)$ denotes the joint probability density of observing a source with unlensed radius~$\beta$ with comoving distance~$\chi_*$, then the \emph{effective reduced moment} of order $n$ is defined as
\begin{equation}\label{eq:effective_moment_def}
\mu\e{n}\h{eff}(\vect{\alpha})
\define \int_0^{\chi\e{H}} \dd\chi_* \, \dd\beta \; p(\beta, \chi_*) \, \mu_n(\chi_*, \vect{\alpha}) \ ,
\end{equation}
where $\chi\e{H}$ is the comoving radius of the particle horizon. For simplicity, we can consider that the intrinsic physical radius~$r$ of a source is independent of its distance from the observer. For a source at $\chi_*$, comoving with the cosmological background, we have $r=f_K(\chi_*)\beta/(1+z_*)$, so that
\begin{align}
p(\beta, \chi_*)
&=  p_\beta(\beta|\chi_*) p_\chi(\chi_*)\\
&= \frac{f_K(\chi_*)}{1+z_*} \, p_r\pac{ \frac{f_K(\chi_*)\beta}{1+z_*}} p_\chi(\chi_*) \ ,
\label{eq:joint_PDF}
\end{align}
where $p_r$ is the probability density function of the intrinsic radius of the sources.

Inserting the expression~\eqref{eq:moment_fixed_z} of $\mu_n$ into Eq.~\eqref{eq:effective_moment_def}, and inverting integration order as $\int_0^{\chi\e{H}}\dd\chi_*\int_0^{\chi_*} \dd\chi = \int_0^{\chi\e{H}}\dd\chi\int_\chi^{\chi\e{H}}\dd\chi_*$, we finally find
\begin{multline}\label{eq:effective_convergence_result}
\mu_n\h{eff}(\vect{\alpha})
= 4\pi G \bar{\rho}_0 \int_0^\infty \dd\beta \int_0^{\chi\e{H}} \dd\chi \, (1+z) f_K(\chi) \\
	\times q(\beta, \chi) \, (\kernel_n*\delta)(\eta_0-\chi, \chi, \vect{\alpha}) \ ,
\end{multline}
with the weighting function
\begin{equation}
q(\beta, \chi)
\define \int_\chi^{\chi\e{H}}
			\dd\chi_* \;  p(\beta, \chi_*) \, \frac{f_K(\chi_*-\chi)}{f_K(\chi_*)} \ ,
\end{equation}
which generalizes to any moment the results that were obtained in \FLUshear for convergence and shear.

\subsection{Two-point correlations}
\label{subsec:two-point_correlations}

We now turn to the heart of this section, which is the definition and calculation of the two-point correlation functions of the image moments. Let $\vect{\alpha}_1$ and $\vect{\alpha}_2$ be two arbitrary directions in the sky, and suppose that we want to correlate the $n_1$th moment of an image observed at $\vect{\alpha}_1$ with the $n_2$th moment of an image at $\vect{\alpha}_2$. Just like for shear, two different correlation functions can be constructed. Call $\phi_{\vect{\alpha}}$ the polar angle of the separation vector~$\vect{\alpha}\define \vect{\alpha}_1 - \vect{\alpha}_2$ between the two lines of sight; then consider a rotated version of the effective moments,
\begin{equation}
\tilde{\mu}_n \define \mu\h{eff}_n \ex{-\i n \phi_{\vect{\alpha}}} \ .
\end{equation}
We define the two correlation functions as
\begin{align}
\xi_{n_1 n_2}^+(\vect{\alpha})
&\define \ev{\tilde{\mu}_{n_1}(\vect{\alpha}_1) \tilde{\mu}_{n_2}^*(\vect{\alpha}_2)} \\
&= \ex{-\i (n_1-n_2)\phi_{\vect{\alpha}}}
		\ev{\mu_{n_1}\h{eff}(\vect{\alpha}_1) \pac{\mu_{n_2}\h{eff}(\vect{\alpha}_2)}^*} \ ,
\label{eq:xi_plus_def}
\end{align}
and
\begin{align}
\xi_{n_1 n_2}^-(\vect{\alpha})
&\define \ev{\tilde{\mu}_{n_1}(\vect{\alpha}_1) \tilde{\mu}_{n_2}(\vect{\alpha}_2)} \\
&= \ex{-\i (n_1+n_2)\phi_{\vect{\alpha}}}
		\ev{\mu_{n_1}\h{eff}(\vect{\alpha}_1) \mu_{n_2}\h{eff}(\vect{\alpha}_2)} \ ,
\label{eq:xi_minus_def}
\end{align}
where $\ev{\ldots}$ denotes ensemble averaging. The names~$\xi^\pm_{n_1 n_2}$ have been chosen by analogy with cosmic shear: for $n_1=n_2=2$, we indeed recover the standard correlation functions $\xi_+$ and $\xi_-$ of weak lensing.

The detailed calculation of $\xi_{n_1 n_2}^\pm$ is given in Appendix~\ref{app:calculation_correlation_functions}, but the main steps can be summarized as follows. First insert the expression~\eqref{eq:effective_convergence_result} of $\mu_n\h{eff}$ into Eqs.~\eqref{eq:xi_plus_def}, \eqref{eq:xi_minus_def}; in Limber's approximation, the main quantities to be calculated are then
\begin{align}
\ev{ \pa{\kernel_{n_1} *\delta} (\eta, \chi, \vect{\alpha}_1) 
		\times \pa{\kernel_{n_2} *\delta}(\eta, \chi, \vect{\alpha}_2) } \ , \\
\ev{ \pa{\kernel_{n_1} *\delta} (\eta, \chi, \vect{\alpha}_1) 
		\times \pa[1]{\kernel_{n_2}^* *\delta}(\eta, \chi, \vect{\alpha}_2) } \ .
\end{align}
Second, introduce the Fourier transform of $\delta$ and the associated matter power spectrum
\begin{equation}
\ev{ \delta(\eta, \vect{k}_1) \delta(\eta, \vect{k}_2) } 
= (2\pi)^3 \delta\e{D}(\vect{k}_1+\vect{k}_2) \, P_\delta(\eta, k_1) \ ,
\end{equation}
where $\delta\e{D}$ denotes the Dirac distribution. Third, integrate over $\vect{\lambda}_1, \vect{\lambda}_2$, as included in the convolution products, which yields Bessel functions $J_{n\pm 1}$. Finally, integrate over the azimuthal angle of $\vect{k}$, which generates another Bessel function~$J_{n_1\pm n_2}$. The final result is
\begin{align}
\xi^+_{n_1 n_2}(\vect{\alpha}) &= \frac{1}{2\pi} 
															\int_0^\infty J_{n_2-n_1}(\alpha\ell) P_{n_1 n_2}(\ell) \;\ell\,\dd\ell \ ,\\
\xi^-_{n_1 n_2}(\vect{\alpha}) &= \frac{(-1)^{n_1}}{2\pi} 
															\int_0^\infty J_{n_1+n_2}(\alpha\ell) P_{n_1 n_2}(\ell) \;\ell\,\dd\ell \ ,
\end{align}
where the power spectra~$P_{n_1 n_2}$ are defined by

\vspace*{0.1cm}

\noindent\fbox{
\begin{minipage}{0.95\columnwidth}
\vspace*{-0.4cm}
\begin{multline}\label{eq:moment_power_spectrum}
P_{n_1 n_2}(\ell) 
=
\pa{\frac{3}{2} H_0^2 \Omega\e{m}}^2
\int_0^{\chi\e{H}} \dd\chi \; (1+z)^2 \\
\times \bar{q}_{n_1}(\ell, \chi) \bar{q}_{n_2}(\ell, \chi) \, P_\delta\pac{\eta_0-\chi, \frac{\ell}{f_K(\chi)} } ,
\end{multline}
with
\begin{equation}\label{eq:kernel_moment}
\bar{q}_n
\define \int_0^\infty \dd\beta \; \frac{4 J_n'(\ell\beta)}{\ell\beta}
			 \int_\chi^{\chi\e{H}} \dd\chi_* \; 
				p(\beta, \chi_*) \, \frac{f_K(\chi_*-\chi)}{f_K(\chi_*)} \ ,
\end{equation}
\end{minipage}
}

\vspace*{1mm}

\noindent where we replaced $4\pi G\bar{\rho}_0$ with $3 H_0^2\Omega\e{m}/2$, $H_0$ being today's cosmic expansion rate, and $\Omega\e{m}$ the cosmological parameter associated with matter density. 

For $n_1=n_2=2$, these results are consistent with what is obtained for shear in \FLUshear. Note also that, even though the original definition of the reduced moments $\mu_n$ is only valid for $n\geq 1$, if we compare Eq.~\eqref{eq:moment_power_spectrum} with the results of \FLUshear for convergence, we find
\begin{equation}
\xi_\kappa(\alpha) = \frac{\xi^\pm_{00}(\alpha)}{4} \ ,  \qquad P_\kappa(\ell) = \frac{P_{00}(\ell)}{4} \ ,
\end{equation}
using that $J_0'(x)=-J_1(x)$.

\begin{figure}[h!]
\centering
\includegraphics[width=\columnwidth]{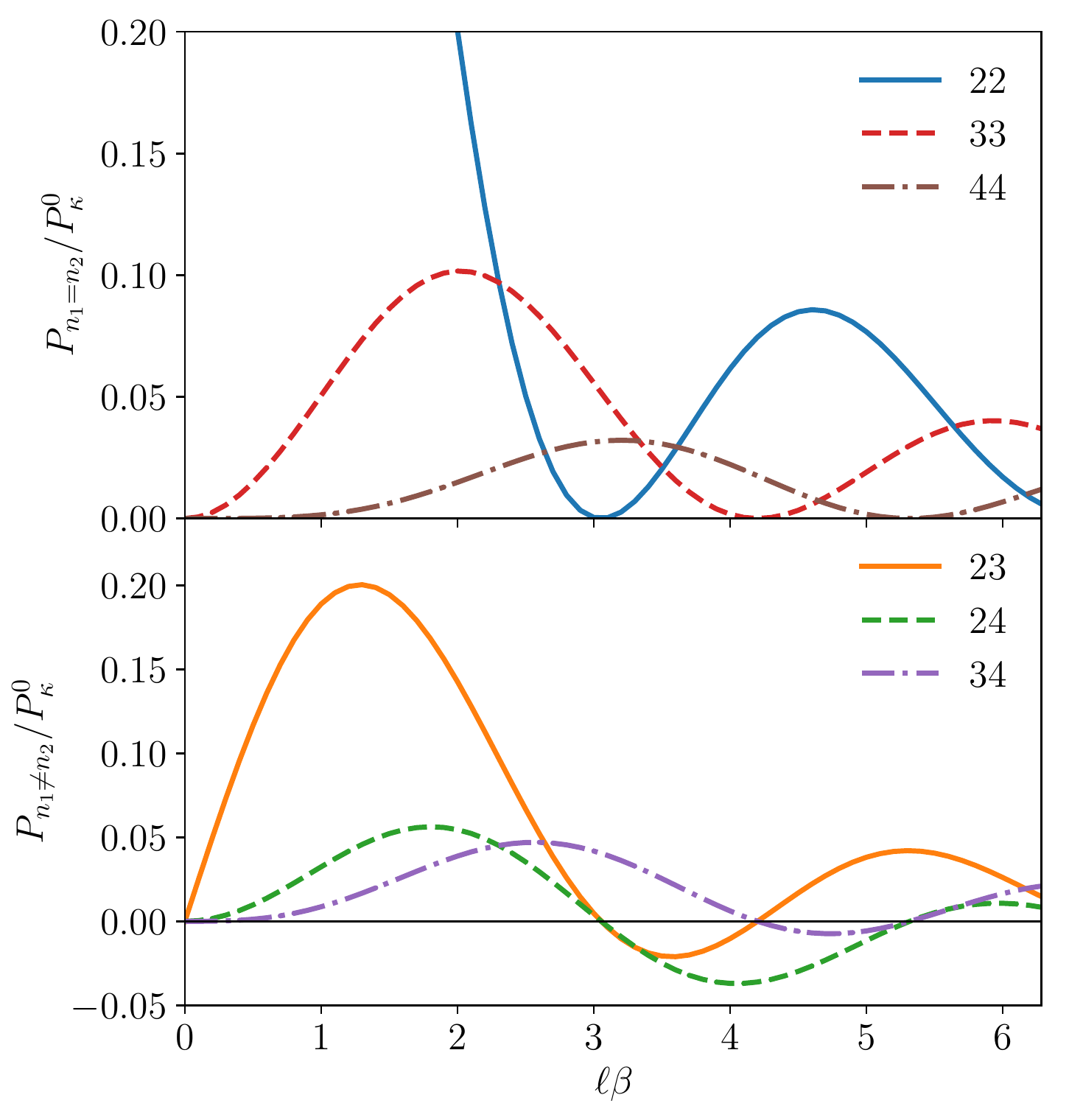}
\caption{Ratio $P_{n_1 n_2}/P_\kappa^0$ between the power spectra of the first reduced moments~$\mu_{n_1}, \mu_{n_2}$ and the standard convergence power spectrum~$P_\kappa^0$ (with infinitesimal sources), for identical circular sources with unlensed radius~$\beta$ located at the same redshift. The top panel shows autocorrelations ($n_1=n_2$); $P_{22}/P_\kappa^0=P_\gamma/P_\kappa^0$ approaches $1$ for $\ell\beta\ll 1$. The bottom panel shows cross correlations ($n_1\not= n_2$).}
\label{fig:damping_functions}
\end{figure}

An instructive special case is when all the sources are identical, and located at the same distance from the observer. Then $p(\beta',\chi_*')=\delta\e{D}(\beta'-\beta)\delta\e{D}(\chi'_*-\chi_*)$, and
\begin{equation}
P_{n_1 n_2}(\ell) = \frac{4 J_{n_1}'(\ell\beta)}{\ell\beta} \,  \frac{4 J_{n_2}'(\ell\beta)}{\ell\beta} \, P_\kappa^0(\ell) \ ,
\end{equation}
where $P^0_\kappa(\ell)$ is the standard convergence power spectrum of convergence with infinitesimal sources. Figure~\ref{fig:damping_functions} illustrates the behavior of $P_{n_1 n_2}(\ell)/P_\kappa^0(\ell)$ for the first $n_1, n_2$. The key piece of information contained in this figure is that, apart from the autocorrelation of shear ($n_1=n_2=2$), all the power spectra vanish when $\beta\rightarrow 0$, and peak for $\ell\sim (\text{a few})/\beta$. The former fact is not surprising: $\beta\rightarrow 0$ corresponds to the infinitesimal-source case, which can only be focused and sheared; in other words, $\mu_{n>2}=0$ in that case, so that the corresponding correlations obviously vanish.

Now consider a more general case where sources are distributed in redshift and apparent radius. For that purpose, we follow the exact same setting as in \FLUshear where the reader can find further details. We consider Milky Way-like galaxies, modeled as perfect disks with physical radius~$R=10\U{kpc}$, and randomly oriented. For simplicity, we still proceed as if these sources were circular, but we allow for their inclination $\iota$ with respect to the line of sight by giving them an apparent radius~$r$ such that $\pi r^2 = \pi R^2|\cos\iota|$. The redshift distribution is taken to be the one\footnote{\href{http://kids.strw.leidenuniv.nl/cosmicshear2016.php}{\tt http://kids.strw.leidenuniv.nl/cosmicshear2016.php}} of the Kilo-Degree Survey (KiDS)~(e.g. Ref.~\cite{Hildebrandt:2016iqg}), in which sources are observed for $z\in[0,0.9]$. Besides, we generate the matter power spectrum~$P_\delta(\eta, k)$ with CAMB\footnote{\href{https://camb.info}{\tt https://camb.info}}, with HALOFIT for nonlinear scales. Cosmological parameters correspond to the \textsl{Planck} 2015 results~\cite{Ade:2015xua}.

\begin{figure}[ht]
\centering
\includegraphics[width=\columnwidth]{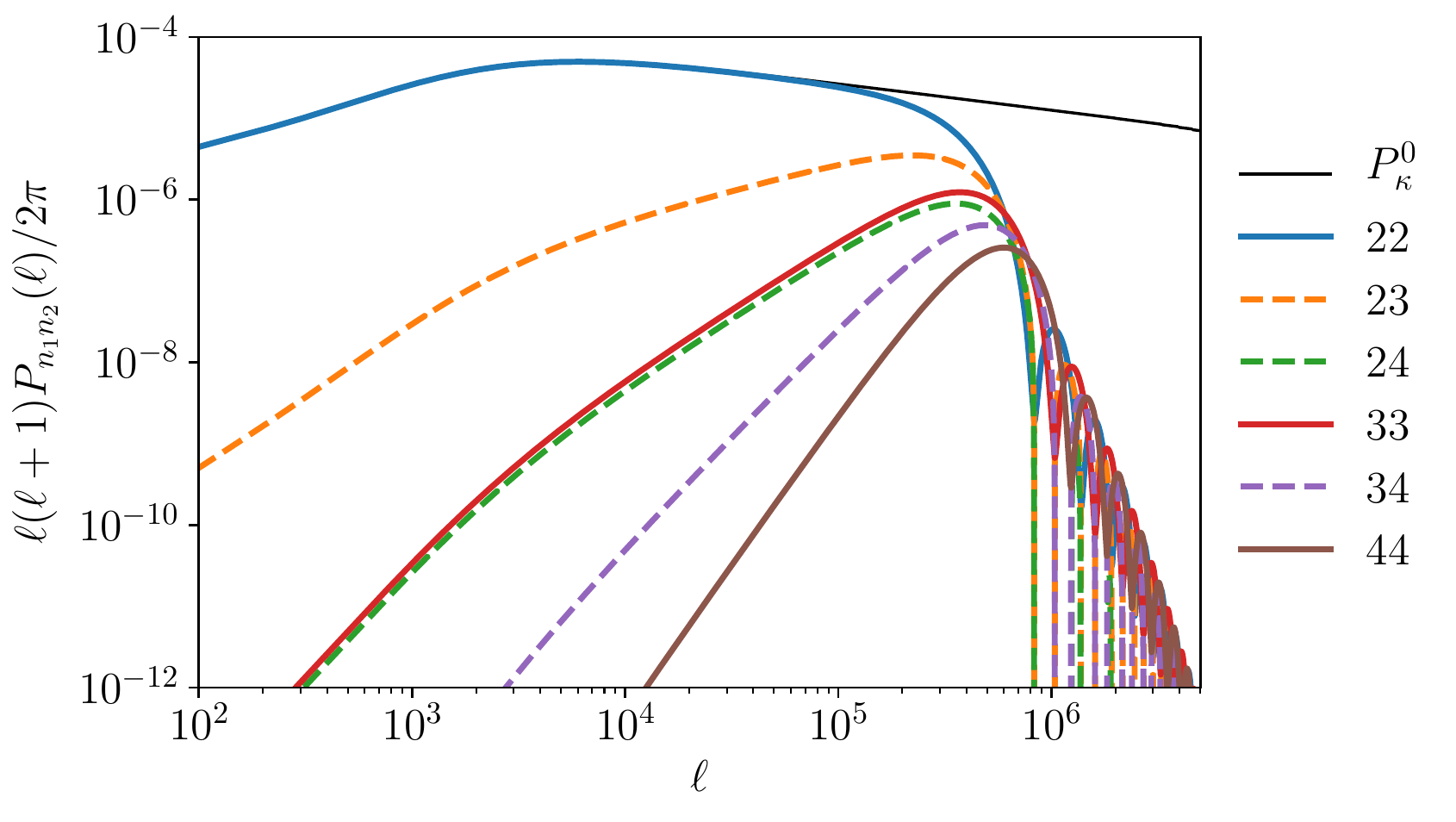}
\caption{Realistic power spectra~$P_{n_1 n_2}(\ell)$ of the first image moments for a KiDS-like survey, with the standard convergence power spectrum~$P_\kappa^0$ for comparison. Autocorrelation spectra are indicated by solid lines, and cross-correlation spectra are indicated by dashed lines; otherwise the order of the curves follows the order of the legend.}
\label{fig:power_spectra}
\end{figure}

The resulting power spectra~$P_{n_1 n_2}(\ell)$, for $n_1, n_2 =2,3,4$ are depicted in Fig.~\ref{fig:power_spectra}, together with $P_\kappa^0$ for comparison. As was already suspected from the simple case of Fig.~\ref{fig:damping_functions}, we see that the power is more localized towards $\ell\sim \beta^{-1}$ as $n_1, n_2$ increase. The amplitudes of the correlations of moments beyond shear only become important when the extended-source corrections to shear ($P_{22}$ compared with $P_\kappa^0$) become significant. This was expected because both effects have the same physical origin, and involve the same characteristic scale---the typical apparent radius of the sources. This scale is extremely small: the typical angular radius of a galaxy at $z\approx 0.5$ is $\beta\sim 1\U{arcsec}$. This essentially corresponds to the maximal angular resolution of an ideal lensing survey---i.e. limited by the number of galaxies that can be observed in the Universe. This resolution remains far beyond the reach of current surveys~\cite{Hildebrandt:2016iqg, Abbott:2017wau}, for which $\alpha\e{min}$ is on the order of a few arcmin, corresponding to a maximum $\ell$ of a few thousands. Note finally that the common behavior of all spectra of Fig.~\ref{fig:power_spectra} at large $\ell$ corresponds to the common asymptotics of the Bessel functions,
\begin{equation}
J_n(x) \approx \frac{\pi}{4}\frac{\cos(x-\pi/4)}{\sqrt{x}}
\qquad \text{for $x\gg 1$.}
\end{equation}

\section{Noncircular sources}
\label{sec:noncircular}

Let us now relax the assumption of circularity of the sources, and investigate how it may change the observed moments of their images. Calculations turn out to be much more challenging, hence, after discussing the most general case in Sec.~\ref{subsec:noncircular_general}, we focus on the case of elliptical sources in Sec.~\ref{subsec:elliptical_sources}, and in particular how it affects shear measurements in Sec.~\ref{subsec:elliptical_shear}. We also propose a perturbative approach in Sec.~\ref{subsec:quasicircular}

\subsection{General case}
\label{subsec:noncircular_general}

Let us go back to the expression~\eqref{eq:reduced_moment_ABCD} of the reduced moments~$\mu_n$, and to a discrete description of lenses. As already mentioned at the end of Sec.~\ref{subsec:reduced_moments_weak_lensing}, the difficulty consists in calculating the four integrals $A_n, B_n, C_n, D_n$, and in particular the first three
\begin{align}
A_n(\cplx{\lambda})
&\define \frac{1}{2\pi \i}
\ointctrclockwise_\source \frac{\cplx{\beta}^n \, \dd\cplx{\beta}}{\cplx{\beta}-\cplx{\lambda}} 
\ , \\
B_n(\cplx{\lambda})
&\define \frac{1}{2\pi \i}
\ointctrclockwise_\source \frac{(\cplx{\beta}^*)^n \, \dd\cplx{\beta}}{\cplx{\beta}-\cplx{\lambda}}
\ , \\
C_n(\cplx{\lambda})
&\define \frac{1}{2\pi \i}
\ointctrclockwise_\source \frac{\beta^n \, \dd\cplx{\beta}}{\cplx{\beta}-\cplx{\lambda}}
\ .
\end{align}
From the residue theorem, $A_n(\cplx{\lambda})=\cplx{\lambda}^n$ if $\cplx{\lambda}\in\interior\source$ and $0$ otherwise, regardless of the shape of the source, but such a direct integration is impossible for $B_n, C_n$, whose integrands are generally not $\mathbb{C}$-differentiable; this is due to the presence of $\cplx{\beta}^*$ in both of them.

In the case of circular sources, this issue was circumvented using that, for any $\cplx{\beta}$ on a circle, $\cplx{\beta}^*\propto\cplx{\beta}^{-1}$. For noncircular sources, however, this trick cannot be applied. Nevertheless, it is still theoretically possible to map this general problem back to the circular case. The Riemann mapping theorem~\cite{Riemann} states that, whatever the shape of $\source$, there exists a biholomorphic\footnote{A biholomorphic function is a one-to-one and onto holomorphic function whose inverse function~$f^{-1}$ is also holomorphic.} function
\begin{equation}
f:\interior\cir_1\rightarrow\interior\source
\end{equation}
which maps the interior of the unit circle~$\cir_1$ to the interior of the source~$\source$. Consider such a map~$f$, and assume without loss of generality that it preserves the orientation of the contours; then we can use it to change variables in~$B_n, C_n$; for instance, $B_n$ becomes
\begin{equation}
B_n = \ointctrclockwise_{\cir_1} \frac{[f(w)^*]^n f'(w)}{f(w)-\cplx{\lambda}} \; \dd w \ .
\end{equation}
Since $f$ is holomorphic, it admits the series expansion
\begin{equation}
f(w) = \sum_{p=0}^\infty f_p \, w^p \ .
\end{equation}
Let us call $f^*$ the function whose coefficients of the Taylor expansion are $f^*_p$. Then, since $w\in\cir_1$, we can use $w^*=1/w$ and the integrals $B_n, C_n$ finally become
\begin{align}
\label{eq:B_n_z}
B_n &= \frac{1}{2\pi \i}\ointctrclockwise_{\cir_1} \frac{[f^*(1/w)]^n f'(w)}{f(w)-\cplx{\lambda}} \; \dd w \ ,\\
\label{eq:C_n_z}
C_n &= \frac{1}{2\pi \i}\ointctrclockwise_{\cir_1} \frac{[f^*(1/w)]^{n/2} [f(w)]^{n/2} f'(w)}{f(w)-\cplx{\lambda}} \; \dd w \ .
\end{align}

The integrands of Eqs.~\eqref{eq:B_n_z} and \eqref{eq:C_n_z} are now explicitly~$\mathbb{C}$-differentiable, and the residue theorem can be applied. Of course, the real difficulty consists in finding the map~$f$, whose construction is not specified by the Riemann mapping theorem. In practice, a possible strategy can consist of experimenting the other way around: starting from known biholomorphic functions~$f$, and generating sources from them.

\subsection{Elliptical sources}
\label{subsec:elliptical_sources}

\begin{figure}[h!]
\centering
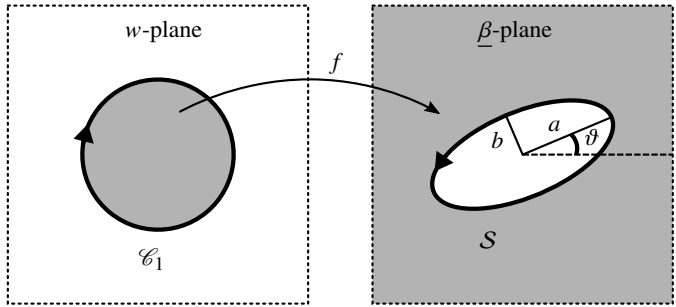
\caption{Complex map $f$ from the interior of the unit disk to the exterior of an ellipse.}
\label{fig:complex_map}
\end{figure}

Most sources used in weak-lensing surveys are elliptical; it is thus relevant to specify the rest of the analysis to ellipses. An example of Riemann map~$f$ from the unit disk to an ellipse can be found in Ref.~\cite{Kanas2006}, but it involves elliptical functions which are not quite easy to handle. However, for the problem at hand, we can use a slightly more convenient method by defining the mapping as follows:
\begin{equation}
f:\left\{
\begin{aligned}
\interior\cir_1 &\longrightarrow \exterior\source \\
w &\longmapsto \cplx{\beta} = \frac{1}{2} \beta_0 \ex{\i\vartheta}\pa{\frac{w}{\ex{\xi}}+ \frac{\ex{\xi}}{w} }
\end{aligned}\right.
\end{equation}
which maps the interior of the unit disk to the \emph{exterior} of the ellipse with semimajor axis~$a=\beta_0\cosh\xi$, inclined with an angle~$\vartheta$ with respect to the real axis, and semiminor axis $b=\beta_0\sinh\xi$ (see Fig.~\ref{fig:complex_map}). This source has complex ellipticity
\begin{equation}
E\e{S} = \frac{a^2-b^2}{a^2+b^2} \, \ex{2\i\vartheta} = \frac{\ex{2\i\vartheta}}{\cosh 2\xi} \ , 
\end{equation}
and area
\begin{equation}
\Omega\e{S} = \pi a b = \frac{\pi}{2} \, \beta_0^2 \sinh 2\xi \ .
\end{equation}
The circular limit is obtained for $\xi\rightarrow\infty$, $\beta_0\rightarrow 0$, while $\beta_0\ex{\xi}\rightarrow \beta\e{c}$, where $\beta\e{c}$ is the radius of the limit circle. It is convenient to introduce the notation
\begin{equation}
\cplx{\beta}_0 \define \beta_0 \ex{\i\vartheta} \ .
\end{equation}
Note that, since $\sqrt{a^2-b^2}=\beta_0$, the complex numbers $\pm\cplx{\beta}_0$ represent the positions of the ellipse's two foci.

The function $f:\interior\cir_1\rightarrow\exterior\source$ is one-to-one and onto. It maps the unit circle to the contour of the source, $f(\cir_1)=\source$, but flipping orientation: if $w$ runs clockwise around $\cir_1$, then $f(w)$ runs anticlockwise around $\source$. The inverse of $f$ is
\begin{equation}
f^{-1}(\cplx{\beta}) =
w = \ex{\xi} \frac{\cplx{\beta}}{\cplx{\beta}_0}
									\pac{ 1 - \sqrt{1 - \pa{\frac{\cplx{\beta}_0}{\cplx{\beta}}}^2} } \ .
\end{equation}

Substituting $f$ in Eqs.~\eqref{eq:B_n_z} and \eqref{eq:C_n_z} yields
\begin{equation}
B_n
= \frac{1}{2\pi \i}\ointclockwise_{\cir_1} b_n(w) \; \dd w \ ,
\quad
C_n
= \frac{1}{2\pi \i}\ointclockwise_{\cir_1} c_n(w) \; \dd w \ ,
\end{equation}
with
\begin{multline}
b_n(w)
= \pa{\frac{\cplx{\beta}_0}{2}}^n
\frac{(\ex{-\xi} w^{-1} + \ex{\xi}w)^n}
		{\ex{-\xi}w + \ex{\xi}w^{-1} - 2\cplx{\lambda}/\cplx{\beta}_0} \\
		\times \frac{\ex{-\xi}w - \ex{\xi} w^{-1}}{w} \ ,
\end{multline}
and
\begin{multline}
c_n(w)
= \pa{\frac{\beta_0}{2}}^n
\frac{[ (\ex{-\xi} w^{-1} + \ex{\xi} w) (\ex{-\xi} w + \ex{\xi} w^{-1}) ]^{\frac{n}{2}}}
		{\ex{-\xi} w + \ex{\xi} w^{-1} - 2\cplx{\lambda}/\cplx{\beta}_0} \\
\times \frac{\ex{-\xi}w - \ex{\xi} w^{-1}}{w} \ .
\end{multline}
Both integrands $b_n, c_n$ have a pole of order $n+1$ at $w=0$. They also have poles for the solutions of $\ex{-\xi}w + \ex{\xi}w^{-1} - 2\cplx{\lambda}/\cplx{\beta}_0=0$, i.e. for $f(w)=\cplx{\lambda}$. Since the only relevant residues are associated to poles located \emph{inside} $\cir_1$, we are interested in solutions of the equation $f(w)=\cplx{\lambda}$ for $w\in\interior\cir_1$. Two cases must be considered:
\begin{enumerate}
\item $\cplx{\lambda}\in\exterior\source$. In this case, since $f:\interior\cir_1\rightarrow\exterior\source$ is one to one and onto, there is one and only one solution to this equation: $w_\lambda\define f^{-1}(\cplx{\lambda})$.
\item $\cplx{\lambda}\in\interior\source$. In that case, there is no solution to $f(w)=\cplx{\lambda}$ within $\interior\cir_1$, because $\cplx{\lambda}\not\in f(\interior\cir_1)$.
\end{enumerate}
Therefore,
\begin{equation}
B_n(\cplx{\lambda}) =
\begin{cases}
-\res_0 b_n & \text{if } \cplx{\lambda} \in \interior\source \ , \\
-\pa{ \res_0 b_n + \res_{w_\lambda} b_n } & \text{if } \cplx{\lambda} \in \exterior\source \ ,
\end{cases}
\end{equation}
and similarly for $C_n(\cplx{\lambda})$; the minus sign before the residues comes from the clockwise orientation of the integration.

The residues at $w_\lambda$ are quite easily calculated. Consider for instance the case of $b_n(w)$; as $w$ approaches $w_\lambda$, we have
\begin{equation}
b_n(w)
= \frac{[f^*(1/w)]^n f'(w)}{f(w)-\cplx{\lambda}}
\sim
\frac{[f^*(1/w_\lambda)]^n}{w-w_\lambda} \ ,
\end{equation}
since $f^*$ is generically regular at $w_\lambda$; whence $\res_{w_\lens}b_n = [f^*(1/w_\lambda)]^n$. With a similar reasoning we find $\res_{w_{\lambda}} c_n = [f(w_\lambda) \, f^*(1/w_\lambda)]^{\frac{n}{2}}$. Replacing $w_\lambda$ with its expression,
\begin{align}
\res_{w_\lambda} b_n
&= \ex{-2\i n\vartheta} \pa{ \cplx{\lambda} \cosh 2\xi - \sqrt{\cplx{\lambda}^2 - \cplx{\beta}_0^2} \, \sinh 2\xi }^n \ ,\\
\res_{w_\lambda} c_n
&= \cplx{\lambda}^{\frac{n}{2}}
	\ex{-\i n\vartheta} \pa{ \cplx{\lambda} \cosh 2\xi - \sqrt{\cplx{\lambda}^2 - \cplx{\beta}_0^2} \, \sinh 2\xi }^{\frac{n}{2}} \ .
\end{align}

As for the pole at $w=0$, since its order is $n+1$, the corresponding residue can be computed with the formula
\begin{equation}
\res_0 b_n = \frac{1}{n!} \left.\ddf[n]{}{w} \pac{ w^{n+1}b_n(w) }\right|_{w=0} \ ,
\end{equation}
and similarly for $c_n$. Although $b_n, c_n$ are rational functions, we do not believe that there exists any simple formula for this derivative for an arbitrary $n$. Nevertheless, it is straightforward to compute it once $n$ has been specified.

\subsection{Corrections to shear}
\label{subsec:elliptical_shear}

In the remainder of this section, we focus on the important case of the complex ellipticity~$E=2\mu_2$. With $n=2$, we can explicitly calculate the last residues, and we get
\begin{align}
\res_0 b_2 &= - (\cplx{\beta}_0^*)^2 \ex{-2\xi} \pac{ \sinh 2\xi + \ex{-2\xi}\pa{\frac{\cplx{\lambda}}{\cplx{\beta}_0}}^2} \ , \\
\res_0 c_2 &= -\beta_0^2 \pac{\frac{1}{2}\sinh 2\xi + \ex{-2\xi}\pa{\frac{\cplx{\lambda}}{\cplx{\beta}_0}}^2} \ .
\end{align}
This ends the computation of the complex integrals involved in the expression~\eqref{eq:reduced_moment_result} of $\mu_2$. The last ingredient is the denominator
\begin{equation}
D_2 \define \frac{1}{2\pi}\int_0^{2\pi} \beta^4(\ph) \; \dd\ph
= \frac{\pi}{4} \, \beta_0^2\sinh 4\xi = \frac{2}{\pi} \frac{\Omega\e{S}^2}{\sqrt{1-|E\e{S}|^2}} \ ,
\end{equation}
which only depends on the shape of the source.

Putting everything together, we find that the ellipticity of the image~$E=2\mu_2$ is related to the ellipticity of the source~$E\e{S}$ by
\begin{equation}
E = E\e{S} \pac{ 1-2\Re(\gamma E\e{S}^*) } + 2\gamma
\end{equation}
which, surprisingly enough, has exactly the same form as in the case of infinitesimal sources---see e.g. Ref.~\cite{1995A&A...294..411S}. What changes is the actual expression of the observed shear~$\gamma$. Like for circular sources, $\gamma$ can be decomposed into a part due to exterior lenses and a contribution of interior lenses:

\vspace*{2mm}

\noindent
\fbox{
\begin{minipage}{0.95\columnwidth}
\vspace*{-4mm}
\begin{equation}
\gamma = \gamma\e{int} + \gamma\e{ext} \ ,
\end{equation}
where, on the one hand,
\begin{align}
\gamma\e{ext}
&= - \sum_{k\in\exterior\source} \pa{\frac{\eps_k}{\cplx{\lambda}_k^*}}^2
														F\pa{\frac{\cplx{\beta}_0^*}{\cplx{\lambda}_k^*}} \ , \\
\text{with} \quad F(z) &\define \frac{8}{z^4} \pa{ 1 - \frac{z^2}{2} - \sqrt{1-z^2} } ,
\end{align}
and, on the other hand,
\begin{multline}
\gamma\e{int}
= \sum_{k\in\interior\source} \eps_k^2 
	\Bigg[ -\frac{2\pi}{\Omega\e{S}} \, \ex{-2\xi+2\i\vartheta}
				+ \sqrt{1-|E\e{S}|^2}\pa{ \frac{\pi \cplx{\lambda}_k}{\Omega\e{S}} }^2 \\
				+ \ex{-4\xi} \sqrt{1-|E\e{S}|^2}\pa{ \frac{\pi \cplx{\lambda}_k^* \ex{2\i\vartheta}}{\Omega\e{S}} }^2
	\Bigg] \ .
\end{multline}
\end{minipage}
}

\vspace*{2mm}

The circular case is recovered for $|E\e{S}|\rightarrow 0$, $\xi\rightarrow\infty$, and $\beta_0=0$. In that regime, we find
\begin{align}
\gamma\e{ext} &\rightarrow - \sum_{k\in\exterior\source} \pa{\frac{\eps_k}{\cplx{\lambda}_k^*}}^2 \\
\gamma\e{int} &\rightarrow \sum_{k\in\interior\source} \pa{\frac{\pi \eps_k \cplx{\lambda}_k}{\Omega\e{S}}}^2 \ ,
\end{align}
where we used $F(z)\rightarrow 1$ for $z\rightarrow 0$, which indeed matches the expression~\eqref{eq:reduced_moment_circular_source_result} of $\mu_n$ for $n=2$. Corrections due to the ellipticity of the source can only be important for a sufficiently extended source. This is obvious for $\gamma\e{int}$, which only exists if the source is extended, while for $\gamma\e{ext}$ it is due to the fact that corrections are controlled by $\beta_0$.

\begin{figure}[ht]
\centering
\includegraphics[width=0.85\columnwidth]{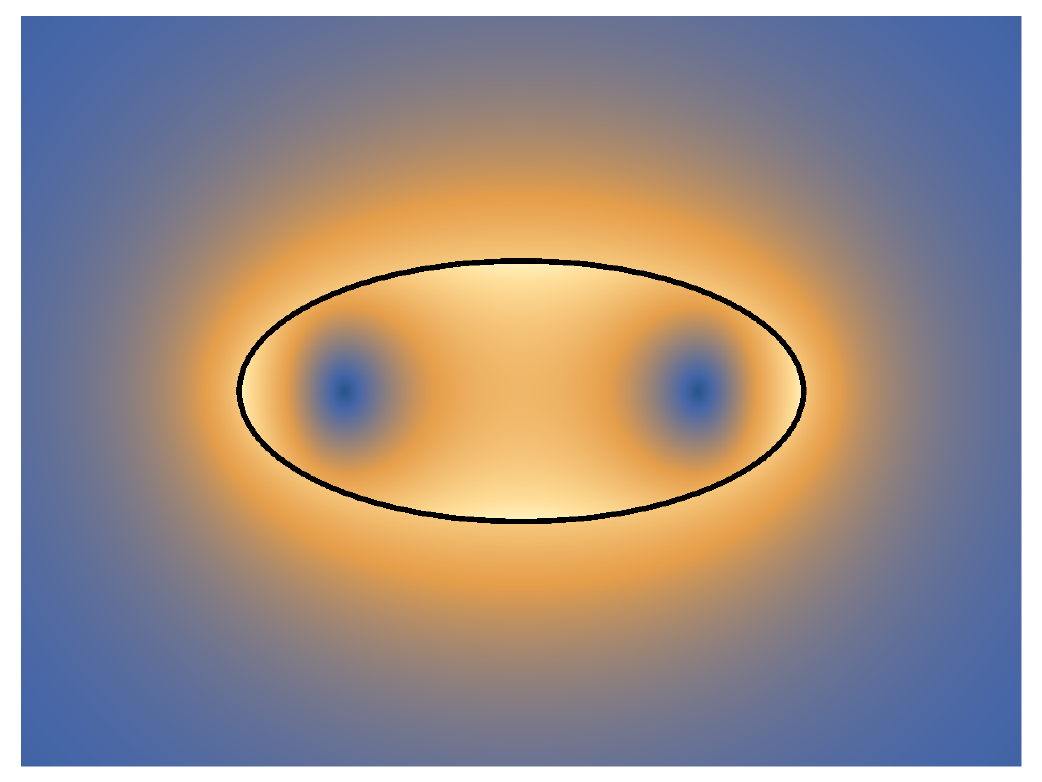}
\hfill
\includegraphics[height=5.4cm]{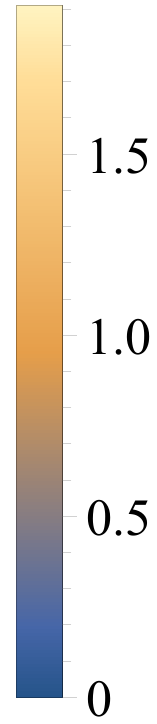}
\caption{Absolute value of the shear~$|\gamma|$ of an elliptical source caused by a single lens, depending on the position of the lens with respect to the source (black line). The color scale indicates the value of $|\gamma|$ in units of the squared Einstein radius~$\eps^2$ of the lens.}
\label{fig:shear_ellipse}
\end{figure}

Figure \ref{fig:shear_ellipse} shows the absolute value~$|\gamma|$ of the shear due to a single lens, depending on the position~$\cplx{\lambda}$ of the lens. As expected, $|\gamma|$ is larger if the lens is closer to the source's contour. Note that $\gamma=0$ on two symmetric points $\cplx{\lambda}=\pm\cplx{\beta}_0/\sqrt{2}$ on the major axis of $\source$. This is where the orientation of $\gamma$ flips: close to the center, an interior lens tends to \emph{reduce} the ellipticity of the source, as seen from the first term~$\propto-\ex{2\i\vartheta}$ of $\gamma\e{int}$; this is an important difference with the circular case, and it is due to the fact that a lens stronger repels the points which are located closer to it. On the contrary, lenses located closer to the foci tend to \emph{enhance} the ellipticity of the source.

For an exterior lens, the first correction with respect to the circular case can be obtained by expanding the function $F$ around zero,
\begin{equation}
F(z) = 1 + \frac{z^2}{2} + \mathcal{O}(z^4) \ ,
\end{equation}
thus, if $\gamma\e{ext}^\circ$ denotes the shear due to a lens at $\cplx{\lambda}=\lambda\ex{\i\phi}$ acting on a circular source, then
\begin{equation}
\frac{|\gamma\e{ext}| - |\gamma\e{ext}^\circ|}{|\gamma\e{ext}^\circ|}
=
\pa{\frac{\beta_0}{\lambda}}^2 \cos 2(\phi-\vartheta) + \ldots
\end{equation}
so that shear is enhanced if the exterior lens is mostly aligned with the major axis of the source, and reduced if it is mostly aligned with its minor axis. This is due to the fact that tidal forces increase as the distance separating two points within the light beam increases.

In a cosmological context, the coupling between shear and intrinsic ellipticity affects the shear power spectrum by changing the expression of the kernel~$\kernel_2(\vect{\lambda})$ involved in Eq.~\eqref{eq:effective_convergence_result}. Its new expression reads~$\kernel_2 = \kernel_2\h{int} + \kernel_2\h{ext}$, with
\begin{multline}
\kernel_2\h{int}(\vect{\lambda})
= -2 \ex{-2\xi+2\i\vartheta} \\
	+ \frac{\pi\lambda^2}{\Omega\e{S}}
		\sqrt{1-|E\e{S}|^2} \pac{ \ex{2\i\phi} + \ex{-4\xi+4\i(\vartheta-\phi)} }
\end{multline}
if $\vect{\lambda}$ lies inside the ellipse $\source$, and zero otherwise, whereas
\begin{equation}
\kernel_2\h{ext}(\vect{\lambda})
= - \frac{\Omega\e{S} \ex{2\i\phi}}{\pi \lambda^2} \, F\pac{ \frac{\beta_0}{\lambda} \, \ex{\i(\phi-\vartheta)} }
\end{equation}
if $\vect{\lambda}$ lies outside the ellipse $\source$, and zero otherwise. These new kernels, combined with the fact that integration must be performed inside and outside an ellipse, instead of inside and outside a circle, makes the calculation of the shear correlation functions more involved.

\subsection{Quasicircular sources}
\label{subsec:quasicircular}

Another way to characterize the effect of noncircularity, which also highlights the entanglement between the intrinsic shape of a source with its lensing distortions, consists in performing a perturbative expansion about the circular case. Let us consider
\begin{equation}
\beta(\varphi) = \bar{\beta} \pac{ 1+\Delta(\ph) } ,
\end{equation}
where $\bar{\beta}$ represents the mean radius of the source, and $|\Delta| \ll 1$ is a real function, with
\begin{equation}
\int_0^{2\pi} \Delta(\ph) \; \dd\ph = 0 \ .
\end{equation}
Recall the general expression~\eqref{eq:reduced_moment_ABCD} of the reduced moments in weak lensing,
\begin{multline}\label{eq:reduced_moment_ABCD_new}
\mu_n
= \paac{
			1 - \frac{n+2}{D_n} \, \sum_k \eps_k^2 \Re\pac{C_n(\cplx{\lambda}_k)}
			} \mu_n\h{S} \\
		+ \frac{1}{D_n} \paac{ \sum_k \eps_k^2 A_n(\cplx{\lambda}_k) 
											+ \sum_k \eps_k^2 \pac{B_n(\cplx{\lambda}_k)}^* } ,
\end{multline}
where the integrals $A_n, B_n, C_n, D_n$ are given by Eqs.~\eqref{eq:A_n}-\eqref{eq:D_n}. Since $\Delta$ has zero mean,
\begin{align}
D_n &= \bar{\beta}^{n+2} + \mathcal{O}(\Delta^2) \ ,\\
\mu_n\h{S} &= \frac{2}{\bar{\beta}}\int_0^{2\pi} \Delta(\varphi) \,\ex{\i n\ph} \; \dd\ph + \mathcal{O}(\Delta^2) \ .
\end{align}
In what follows, we choose to work at first order in $\Delta$, but keep cross terms~$\mathcal{O}(\Delta\times \delta\theta)$. It implies that $C_n$ only needs to be computed at zeroth order in $\Delta$, i.e. as if the source were circular with radius~$\bar{\beta}$,
\begin{equation}
C^{(0)}_n(\cplx{\lambda})
= \frac{1}{2\pi\i} \ointctrclockwise_{\bar{\source}} \frac{\beta^n \dd\cplx{\beta}}{\cplx{\beta}-\cplx{\lambda}}
=
\begin{cases}
\bar{\beta}^n & \text{if } \cplx{\lambda}\in\interior\source \ ,\\
0 & \text{otherwise.}
\end{cases}
\end{equation}

As already emphasized, $A_n$ is given by \eqref{eq:moment_circular_source_first_integral} whatever the shape of the source. What remains to be determined is thus the expansion of $B_n$ at first order in $\Delta$, $B_n=B_n^{(0)}+B_n^{(1)}$. The zeroth order corresponds to the circular case, given by Eq.~\eqref{eq:B_n_circular}; the first order reads
\begin{multline}\label{eq:B^1}
B^{(1)}_n(\cplx{\lambda}) =
\frac{1}{2\pi \i} \int_0^{2\pi} 
	\pac{\ddf{\Delta}{\ph} + \i(n+1)\Delta(\ph)} \frac{\ex{-\i(n-1)\ph}}{\ex{\i\ph}- \cplx{\lambda}/\bar{\beta}} \; \dd\ph \\
   + \frac{1}{2\pi}\int_0^{2\pi}
   	\frac{\Delta(\ph) \, \ex{-\i(n-2)\ph}}{(\ex{\i\varphi} - \cplx{\lambda}/\bar{\beta})^2} \; \dd\ph \ .
\end{multline}
To proceed further, we decompose $\Delta$ in Fourier series as
\begin{equation}
\Delta(\varphi)
=
\sum_{p=1}^\infty \cplx{\Delta}_p \ex{\i p \ph} \ ,
\end{equation}
with $\cplx{\Delta}_p = \cplx{\Delta}_{-p}^*$ since $\Delta$ is a real function. This allows us to compute the integrals of Eq.~\eqref{eq:B^1} and obtain
\begin{equation}
B_n^{(1)}(\cplx{\lambda})
=
\begin{cases}
\displaystyle
\sum_{p=n}^{\infty} (p+1) \cplx{\Delta}_p \pa{ \dfrac{\cplx{\lambda}}{\bar{\beta}} }^{p-n}
& \text{if } \cplx{\lambda}\in\interior\source \ ,\\
\displaystyle
-\sum_{p=-\infty}^{n-1} (p+1) \cplx{\Delta}_p \pa{ \dfrac{\cplx{\lambda}}{\bar{\beta}} }^{p-n}
& \text{if } \cplx{\lambda}\in\exterior\source \ .
\end{cases}
\end{equation}
Gathering all the terms, we conclude that, at first order in $\Delta$ and $\delta\theta$,
\begin{align}
\mu_n
&=
(1 - \kappa ) \, \mu_n\h{S} + \mu_n^\circ  \nonumber \\
&\quad+ \sum_{k\in\interior\source} \pa{\frac{\eps_k}{\bar{\beta}}}^2
		\sum_{p=n+1}^\infty (p+1) \cplx{\Delta}_p^* \pa{ \frac{\cplx{\lambda}_k^*}{\bar{\beta}} }^{p-n} \nonumber \\
&\quad- \sum_{k\in\exterior\source} \pa{\frac{\eps_k}{\bar{\beta}}}^2
		\sum_{p=-\infty}^{n-1} (p+1) \cplx{\Delta}_p^* \pa{ \frac{\cplx{\lambda}_k^*}{\bar{\beta}} }^{p-n} \ ,
\end{align}
where $\mu^{\circ}_n$ corresponds to the reduced moments in the circular case, and is given by Eq.~\eqref{eq:reduced_moment_circular_source_result}. In the above equation, the first term is just the magnification of the intrinsic reduced moment; the second term is the reduced moment generated by lensing on a circular source; the last two terms arise from the coupling between the intrinsic and lensing moments.

\section{Conclusion}
\label{sec:conclusion}

In this article, we have seen how the weak lensing distortions of an extended source can be described by successive moments beyond shear (see Fig.~\ref{fig:Fourier_modes}). We developed a simple and elegant formalism, based on complex analysis, to calculate those moments, and applied it to a realistic cosmological model. As a rule of thumb, for circular sources, the power spectrum of the angular correlation function between the moments of order $n_1, n_2 > 1$ reads
\begin{equation}
P_{n_1 n_2}(\ell) \approx \frac{4 J_{n_1}'(\ell\beta)}{\ell\beta} \, \frac{4 J_{n_2}'(\ell\beta)}{\ell\beta} \, P_\kappa^0(\ell) \ ,
\end{equation}
where $J_n$ are Bessel functions, and $\beta$ is the typical angular radius of the sources, while $P_\kappa^0$ denotes the convergence power spectrum in the infinitesimal-source limit. Higher-order Bessel functions tend to be more peaked, so that $P_{n_1 n_2}(\ell)$ gets more and more peaked at $\ell\sim\beta^{-1}$ as $n_1, n_2$ increase. Correlations between high-order moments thus only occur on scales comparable to the source's size. Although the correlation of higher-order distortion modes may seem far from what is currently achievable in astronomy, new type of sources, such as Einstein rings themselves~\cite{Birrer:2017sge}, could make such features observable in the future.

We also have shown that our formalism can be applied to noncircular sources, thanks to a variation on the Riemann mapping theorem, and we illustrated this method to the astrophysically relevant case of elliptic sources. An important conclusion is the entanglement between lensing moments and intrinsic moments; contrary to what happens with infinitesimal sources, where the shear~$\gamma$ is independent from the intrinsic ellipticity of the source, for extended sources this intrinsic ellipticity directly affects the value of shear, and of the other distortion modes. This is reminiscent of the results of Ref.~\cite{Viola:2011ue} about the impact of image ellipticities on flexion measurements. The entanglement grows with the size of the source, and hence becomes significant precisely when other extended-source corrections become important as well. Therefore, noncircularity does not change whether finite-beam effects are significant or not, but it affects their behavior when they are.

This latter conclusion naturally calls for an extension of the analysis of Sec.~\ref{sec:cosmic} to elliptical sources. Indeed, since the correlations of image moments are increasingly sensitive to scales comparable to the beam's size as the order of the moment increases, we expect corrections due to the source's ellipticity to strongly affect high-order moment power spectra. If intrinsic ellipticities are randomly oriented, we expect to recover results close to the circular case on average. However, intrinsic alignments~\cite{Troxel:2014dba, Joachimi:2015mma} might affect this expectation.

An important restriction of the analysis of this article resides in our choice of top-hat weighting function~$W[I(\vect{\theta})]$ in the definition of the image moments. This choice was mathematically very convenient, since it allowed us to convert an initially two-dimensional problem into a one-dimensional problem---the analysis of the image contour. However, as discussed in Sec.~\ref{subsec:comparison_flexion_roulettes}, it removes part of the information contained in the image; in particular, it makes the $\flexionF$-type flexion unobservable. Generalizing the present approach to any weighting function would thus add great value to the understanding of weak gravitational lensing beyond shear.

\section*{Acknowledgements} We thank David Bacon, Cyril Pitrou, and Fabien Lacasa for useful and stimulating discussions. P.F. acknowledges support by the Swiss National Science Foundation. The work of J.-P.U. is made in the ILP LABEX (under reference ANR-10-LABX-63) was supported by French state funds managed by the ANR within the Investissements d'Avenir program under Reference No ANR-11-IDEX-0004-02. J.L.'s work is supported by the National Research Foundation (South Africa).

\onecolumngrid
\appendix
\section{Calculation of the two-point correlation functions of the image moments}
\label{app:calculation_correlation_functions}

We consider here the distortions of circular sources. Let $\vect{\alpha}_1, \vect{\alpha}_2$ be two directions in the (flat) sky, and $\vect{\alpha}\define \vect{\alpha}_1-\vect{\alpha}_2 = \alpha(\cos\phi_{\vect{\alpha}}, \sin\phi_{\vect{\alpha}})$ their separation. The two correlation functions of the $n_1$th and $n_2$th image moments were defined in Sec.~\ref{subsec:two-point_correlations} as
\begin{align}
\xi_{n_1 n_2}^+(\alpha)
&\define \ex{-\i (n_1-n_2)\phi_{\vect{\alpha}}}
		\ev{\mu_{n_1}\h{eff}(\vect{\alpha}_1) \pac{\mu_{n_2}\h{eff}(\vect{\alpha}_2)}^*} \ ,\\
\xi_{n_1 n_2}^-(\alpha)
&\define \ex{-\i (n_1+n_2)\phi_{\vect{\alpha}}}
		\ev{\mu_{n_1}\h{eff}(\vect{\alpha}_1) \mu_{n_2}\h{eff}(\vect{\alpha}_2)} \ .
\end{align}

\subsection{Introducing the matter power spectrum and Limber's approximation}

Let us first consider $\xi^-_{n_1 n_2}$, the calculation of $\xi^+_{n_1 n_2}$ following essentially the same lines. We start by substituting the definition of the effective reduced moments as follows:
\begin{multline}
\ev{\mu_{n_1}\h{eff}(\vect{\alpha}_1) \mu_{n_2}\h{eff}(\vect{\alpha}_2)}
= (4\pi G \bar{\rho}_0)^2
	\int_0^\infty \dd\beta_1 \, \dd\beta_2 \int_0^{\chi\e{H}} \dd\chi_1 \dd\chi_2 \;
	(1+z_1) f_K(\chi_1) q(\beta_1, \chi_1) \, (1+z_2) f_K(\chi_2) q(\beta_2, \chi_2) \\
 	\times \ev{
 						(\kernel_{n_1} * \delta)(\eta_1, \chi_1, \vect{\alpha}_1) 
 						(\kernel_{n_2} * \delta)(\eta_2, \chi_2, \vect{\alpha}_2)
 						}
\end{multline}
where it is understood that $\eta_1=\eta_0-\chi_1$ and $\eta_2=\eta_0-\chi_2$, since the density contrast is evaluated on the (background) light cone of the observer. By making the convolution products explicit, and inserting the Fourier transform of the density contrast, we have
\begin{align}
\ev{(\kernel_{n_1} * \delta)\,(\kernel_{n_2} * \delta)}
&= \int \frac{\dd^2\vect{\lambda}_1}{\pi\beta_1^2} \,  \frac{\dd^2\vect{\lambda}_2}{\pi\beta_2^2} \;
		\kernel_{n_1}(\vect{\lambda}_1) \, \kernel_{n_2}(\vect{\lambda}_2)
		\ev{ \delta(\eta_1, \chi_1, \vect{\alpha}_1+\vect{\lambda}_1) \,
				\delta(\eta_2, \chi_2, \vect{\alpha}_2+\vect{\lambda}_2)	 } \\
&=  \int \frac{\dd^2\vect{\lambda}_1}{\pi\beta_1^2} \,  \frac{\dd^2\vect{\lambda}_2}{\pi\beta_2^2} \,
				\frac{\dd^3\vect{k}_1}{(2\pi)^3} \, \frac{\dd^3\vect{k}_1}{(2\pi)^3} \;
		\ex{\i(\vect{k}_1\cdot\vect{x}_1 + \vect{k}_2\cdot\vect{x}_2)} \,
		\kernel_{n_1}(\vect{\lambda}_1) \, \kernel_{n_2}(\vect{\lambda}_2)
		\ev{ \delta(\eta_1, \vect{k}_1) \,
				\delta(\eta_2, \vect{k}_2)	 } \\
&=  \int \frac{\dd^2\vect{\lambda}_1}{\pi\beta_1^2} \,  \frac{\dd^2\vect{\lambda}_2}{\pi\beta_2^2} \,
				\frac{\dd^3\vect{k}}{(2\pi)^3} \;
		\ex{\i \vect{k}\cdot(\vect{x}_1 - \vect{x}_2)} \,
		\kernel_{n_1}(\vect{\lambda}_1) \, \kernel_{n_2}(\vect{\lambda}_2) \, P_\delta(\eta_1, \eta_2, k) \ ,
\end{align}
where $\vect{x}_1$ is the spatial position corresponding to $\chi_1, \vect{\alpha}_1+\vect{\lambda}_1$, and similarly for $\vect{x}_2$. In the last line, we introduced the power spectrum $P_\delta$ with
\begin{equation}
\ev{\delta(\eta_1, \vect{k}_1) \delta(\eta_2, \vect{k}_2)}
= (2\pi)^2 \delta\e{D}(\vect{k}_1+\vect{k}_2) \, P_\delta(\eta_1, \eta_2, k_1) \ ,
\end{equation}
we also integrated over $\vect{k}_2$, and changed the name of $\vect{k}_1$ to $\vect{k}$. We then apply Limber's approximation:  first split the phase of the complex exponential into a longitudinal part and a transverse part,
\begin{equation}
\vect{k} \cdot (\vect{x}_1-\vect{x}_2)
= k_{||} (\chi_1-\chi_2) 
	+ \vect{k}_\perp \cdot \pac{f_K(\chi_1)(\vect{\alpha}_1+\vect{\theta}_1)
													-f_K(\chi_2)(\vect{\alpha}_2+\vect{\theta}_2)} \ .
\end{equation}
Since the configuration for which most correlations occur is $|\chi_1-\chi_2|\ll \chi_1,\chi_2$ (small angles), the major contribution to the integral is such that $k_{||}\ll k_\perp$; thus, we can approximate~$k\approx k_\perp$ in the matter power spectrum, and integrate over $k_{||}$ to get $2\pi \delta\e{D}(\chi_1-\chi_2)$. We could also have performed this reasoning in normal space, arguing that a change in $|\vect{\alpha}_1-\vect{\alpha_2}|$ produces a much more significant change than a change in $\chi_1-\chi_2$: the correlations are mostly transverse. Calling $\vect{\ell} \define f_K(\chi_1)\vect{k}_\perp$, we therefore get
\begin{equation}
\ev{(\kernel_{n_1} * \delta)\,(\kernel_{n_2} * \delta)}
=
\frac{\delta\e{D}(\chi_1-\chi_2)}{f_K^2(\chi_1)}
\int \frac{\dd^2\vect{\lambda}_1}{\pi \beta_1^2}
		\, \frac{\dd^2\vect{\lambda}_2}{\pi \beta_2^2}
		\, \frac{\dd^2\vect{\ell}}{(2\pi)^2} \;
		\ex{\i\vect{\ell}\cdot (\vect{\alpha}_1-\vect{\alpha}_2) 
				+ \i\vect{\ell}\cdot (\vect{\lambda}_1-\vect{\lambda}_2)}
		\, \kernel_{n_1}(\vect{\lambda}_1) \, \kernel_{n_2}(\vect{\lambda}_2) \,
		P_\delta \pac{\eta_1, \frac{\ell}{f_K(\chi_1)}} \ .
\end{equation}
Inserting the above into the definition of $\xi^-_{n_1 n_2}$, and noticing that the same calculation applies to $\xi^+_{n_1 n_2}$ if one turns $\kernel_{n_2}$ into $\kernel_{n_2}^*$, we can put the correlation functions under the form
\begin{align}
\label{eq:xi_plus_end_first_step}
\xi^+_{n_1 n_2}(\alpha)
&= (4\pi G \bar{\rho}_0)^2
	\int_{\mathbb{R}^2} \frac{\dd^2\vect{\ell}}{(2\pi)^2} \; \ex{\i\vect{\ell}\cdot(\vect{\alpha}_1-\vect{\alpha}_2)}
	\int_0^{\chi\e{H}} \dd\chi \; (1+z)^2 \,
			g_{n_1}(\chi, \vect{\ell},\vect{\alpha}) \, g_{n_2}^*(\chi, \vect{\ell},\vect{\alpha}) \,
			P_\delta \pac{\eta_0-\chi, \frac{\ell}{f_K(\chi)}} \ , \\
\label{eq:xi_minus_end_first_step}
\xi^-_{n_1 n_2}(\alpha)
&= (4\pi G \bar{\rho}_0)^2
	\int_{\mathbb{R}^2} \frac{\dd^2\vect{\ell}}{(2\pi)^2} \; \ex{\i\vect{\ell}\cdot(\vect{\alpha}_1-\vect{\alpha}_2)}
	\int_0^{\chi\e{H}} \dd\chi \; (1+z)^2 \,
			g_{n_1}(\chi, \vect{\ell},\vect{\alpha}) \, g_{n_2}(\chi, -\vect{\ell},\vect{\alpha}) \,
			P_\delta \pac{\eta_0-\chi, \frac{\ell}{f_K(\chi)}} \ ,
\end{align}
with
\begin{equation}
g_n(\chi, \vect{\ell}, \vect{\alpha})
\define
			\int_0^\infty \dd\beta \; q(\beta, \chi)
			\int_{\mathbb{R}^2} \frac{\dd^2\vect{\lambda}}{\pi\beta^2} \; 
					\ex{\i(\vect{\ell}\cdot\vect{\lambda}-n \phi_{\vect{\alpha}})}\,\kernel_n(\vect{\lambda}) \ .
\end{equation}

\subsection{Calculation of $g_n$}

The next step of the calculation consists in performing the integration over $\vect{\lambda}$ in order to get the explicit expression of $g_n(\chi, \vect{\ell},\vect{\alpha})$. Replacing the kernel~$\kernel_n = \kernel_n\h{int} + \kernel_n\h{ext}$ with its expression, and using polar coordinates for both $\vect{\ell}$ and $\vect{\lambda}$, with $\vect{\lambda}=\lambda(\cos\phi, \sin\phi)$ and $\vect{\ell}=\ell(\cos\phi_{\vect{\ell}}, \sin\phi_{\vect{\ell}})$, we have
\begin{align}
\int_{\mathbb{R}^2} \frac{\dd^2\vect{\lambda}}{\pi\beta^2} \; 
					\ex{\i(\vect{\ell}\cdot\vect{\lambda}-n \phi_{\vect{\alpha}})}\,\kernel_n(\vect{\lambda})
&= \pac{
			\int_0^\beta \frac{\lambda \, \dd\lambda}{\beta^2} \pa{\frac{\lambda}{\beta}}^n
			- \int_\beta^\infty \frac{\lambda \, \dd\lambda}{\beta^2} \pa{\frac{\beta}{\lambda}}^n
			}
		\int_0^{2\pi} \frac{\dd\phi}{\pi} \; \ex{\i[\ell\lambda \cos(\phi-\phi_{\vect{\ell}}) - n(\phi - \phi_{\vect{\alpha}})]}
		\\
&= \ex{\i n(\phi_{\vect{\ell}}-\phi_{\vect{\alpha}})}
	\pac{
			\int_0^1 \dd x \; x^{1+n}
			- \int_1^\infty \dd x \; x^{1-n}
			}
		\int_0^{2\pi} \frac{\dd\phi}{\pi} \; \ex{\i(x\ell\beta\cos\psi - n\psi)} \ ,
\end{align}
where we introduced $x\define\lambda/\beta$ and $\psi \define \phi-\phi_{\vect{\ell}}$. The angular integral yields a Bessel function as
\begin{equation}
\int_0^{2\pi} \frac{\dd\phi}{\pi} \; \ex{\i(\ell\beta x\cos\psi - n\psi)}
= \ex{\i n \pi/2} \int_0^{2\pi} \frac{\dd\phi}{\pi} \; \ex{-\i (n\psi - \ell\beta x\sin\psi)}
= 2 \i^n \, J_n(x\ell\beta) \ .
\end{equation}
We then use that for any strictly positive $n$,
\begin{align}
\int_0^1 \dd x \; x^{1+n} J_n(x y) = \frac{J_{n+1}(y)}{y} \\
\int_1^\infty \dd x \; x^{1-n} J_n(x y) = \frac{J_{n-1}(y)}{y} \ ,
\end{align}
to get
\begin{equation}
\int_{\mathbb{R}^2} \frac{\dd^2\vect{\lambda}}{\pi\beta^2} \; 
					\ex{\i(\vect{\ell}\cdot\vect{\lambda}-n \phi_{\vect{\alpha}})}\,\kernel_n(\vect{\lambda})
= 2\i^n \ex{\i n(\phi_{\vect{\ell}}-\phi_{\vect{\alpha}})} \pac{ \frac{J_{n+1}(\ell\beta)-J_{n-1}(\ell\beta)}{\ell\beta} } \ ,
\end{equation}
and finally we use that $J_{n+1}(x)-J_{n-1}(x)=-2 J'_n(x)$ to conclude that
\begin{equation}\label{eq:g_n_result}
g_n(\chi, \vect{\ell}, \vect{\alpha})
=
-\i^n \ex{\i n(\phi_{\vect{\ell}}-\phi_{\vect{\alpha}})} \int_0^\infty \dd\beta \; q(\beta, \chi) 
			\, \frac{4 J'_n(\ell\beta)}{\ell\beta} \ .
\end{equation}

\subsection{Final result}

The last step of the calculation consists in integrating over the angular part of $\vect{\ell}$ in the expressions~\eqref{eq:xi_plus_end_first_step} and \eqref{eq:xi_minus_end_first_step} of $\xi^+_{n_1 n_2}$ and $\xi^-_{n_1 n_2}$. Substituting the expression~\eqref{eq:g_n_result} of $g_n$, we can put both correlation functions under the form
\begin{equation}
\xi^\pm_{n_1 n_2}(\vect{\alpha})
= \frac{1}{2\pi} \int_0^\infty J_\pm(\vect{\alpha}, \vect{\ell}) \, P_{n_1 n_2}(\ell) \; \ell \ \dd\ell \ ,
\end{equation}
where
\begin{align}
P_{n_1 n_2}(\ell) 
&=
\pa{4\pi G \bar{\rho}_0}^2
\int_0^{\chi\e{H}} \dd\chi \; (1+z)^2\, \bar{q}_{n_1}(\ell, \chi) \, \bar{q}_{n_2}(\ell, \chi) \, P_\delta\pac{\eta_0-\chi, \frac{\ell}{f_K(\chi)} } ,\\
\text{with} \quad
\bar{q}_n
&\define \int_0^\infty \dd\beta \; \frac{4 J_n'(\ell\beta)}{\ell\beta}
			\, q(\beta, \chi) \ ,
\end{align}
and
\begin{align}
J_+(\vect{\ell}, \vect{\alpha})
&= \i^{n_1-n_2}
	\int_0^{2\pi} \frac{\dd\phi_{\vect{\ell}}}{2\pi} \; 
			\ex{\i\vect{\ell}\cdot\vect{\alpha}}
			\ex{\i (n_1-n_2)(\phi_{\vect{\ell}}-\phi_{\vect{\alpha}})}
= \int_0^{2\pi} \frac{\dd\psi}{2\pi} \; 
			\ex{-\i[(n_2-n_1)\psi - \ell\alpha\sin\psi]}
= J_{n_2-n_1}(\ell\alpha) \ ,
\\
J_-(\vect{\ell}, \vect{\alpha})
&= (-1)^{n_2} \i^{n_1+n_2}
	\int_0^{2\pi} \frac{\dd\phi_{\vect{\ell}}}{2\pi} \;
			\ex{\i\vect{\ell}\cdot\vect{\alpha}}
			\ex{\i (n_1+n_2)(\phi_{\vect{\ell}}-\phi_{\vect{\alpha}})}
= (-1)^{n_2} \int_0^{2\pi} \frac{\dd\psi}{2\pi} \; 
			\ex{-\i[(-n_2-n_1)\psi - \ell\alpha\sin\psi]}
= (-1)^{n_1} J_{n_1+n_2}(\ell\alpha) \ ,
\end{align}
where in the last equality we used $J_{-n}(x)=(-1)^n J_n(x)$. This ends the derivation of the correlation functions of the image reduced moments.

\twocolumngrid
\bibliography{bibliography_finite_beams_advanced}

\end{document}